\documentclass[aps,prd,showpacs,eqsecnum,notitlepage]{revtex4-1}

\usepackage[centertags]{amsmath}
\usepackage{amssymb}
\usepackage{latexsym}
\usepackage{enumerate}
\usepackage{graphicx}
\usepackage{mathrsfs}
\usepackage{hyperref}
\usepackage{stmaryrd}
\usepackage{empheq}

\newcommand{\bi}{\begin{itemize}}
\newcommand{\ei}{\end{itemize}}

\newcommand{\be}{\begin{equation}}
\newcommand{\ee}{\end{equation}}

\newcommand{\ba}{\begin{array}}
\newcommand{\ea}{\end{array}}

\renewcommand{\l}{\left(}
\renewcommand{\r}{\right)}
\renewcommand{\a}{\alpha}
\renewcommand{\b}{\beta}

\renewcommand{\d}{\delta}

\newcommand{\la}{\lambda}
\renewcommand{\O}{\Omega}
\renewcommand{\o}{\omega}
\renewcommand{\th}{\theta}

\newcommand{\q}{\quad}

\renewcommand{\k}{\kappa}

\newcommand{\pa}{\partial}

\renewcommand{\Bar}[1]{\makebox{$\bar{#1}$}}

\begin{document}

\title{Metric perturbations from eccentric orbits on a Schwarzschild black 
hole: I. Odd-parity Regge-Wheeler to Lorenz gauge transformation and two 
new methods to circumvent the Gibbs phenomenon} 

\author{Seth Hopper}
\email{seth.hopper@aei.mpg.de}
\affiliation{Max-Planck-Institut f\"ur Gravitationsphysik, 
Albert-Einstein-Institut, \\ Am M\"uhlenberg 1, D-14476 Golm, Germany}

\author{Charles R. Evans}
\email{evans@physics.unc.edu}
\affiliation{Department of Physics and Astronomy, University of North 
Carolina, Chapel Hill, North Carolina 27599}

\begin{abstract}

We calculate the odd-parity, radiative ($\ell \ge 2$) parts of the metric 
perturbation in Lorenz gauge caused by a small compact object 
in eccentric orbit about a Schwarzschild black hole.  The Lorenz gauge 
solution is found via gauge transformation from a corresponding one in 
Regge-Wheeler gauge.  Like the Regge-Wheeler gauge solution itself, the 
gauge generator is computed in the frequency domain and transferred to the 
time domain.  The wave equation for the gauge generator has a source with 
a compact, moving delta-function term and a discontinuous non-compact term.  
The former term allows the method of extended homogeneous solutions to be 
applied (which circumvents the Gibbs phenomenon).  The latter has required
the development of new means to use frequency domain methods and yet be 
able to transfer to the time domain while avoiding Gibbs problems.  Two new 
methods are developed to achieve this: a partial annihilator method and a 
method of extended particular solutions.  We detail these methods and show 
their application in calculating the odd-parity gauge generator and Lorenz 
gauge metric perturbations.  A subsequent paper will apply these methods 
to the harder task of computing the even-parity parts of the gauge generator.

\end{abstract}


\maketitle

\section{Introduction}
\label{sec:intro}

The increasing maturity of space-based gravitational wave detector 
concepts~\cite{NASA,ESA} has in part motivated considerable work in the 
last fifteen years on self-consistent calculations of extreme-mass-ratio 
inspirals (EMRIs).  Such a system consists of a small compact object of mass 
$\mu \simeq 1-10 M_{\odot}$ (e.g., neutron star or black hole) moving on a 
decaying orbit about, and ultimately into, a supermassive black hole of 
mass $M \sim 10^5 - 10^9 M_{\odot} \gg \mu$.  Irrespective of when a detector 
might launch, there is also simply considerable theoretical interest in the 
problem of motion of a point mass in a background geometry in general 
relativity, influenced by its own self-force~\cite{Poisson_2011}.

The extreme mass-ratio lends itself to use of black hole perturbation theory.  
In the limit of $\mu \rightarrow 0$ the small mass orbits on a geodesic of
the massive black hole background with constants of motion.  At next order 
the small mass draws up a small perturbation in the metric, which results 
in gravitational radiation fluxing to infinity and down the horizon of the 
massive black hole.  The metric perturbation (MP) also acts back on the small 
body locally (through a self-force), giving rise to dissipative effects that 
cause the orbit to decay and to small conservative corrections to the motion.  
The perturbation problem is singular in several respects~\cite{Pound_2010}, 
with a divergence in the MP at the particle location making 
the motion correction also formally divergent.  A general understanding of 
how to treat the self-force (i.e., regularize it) in an arbitrary spacetime 
was given by Mino, Sasaki, and Tanaka~\cite{MiSaTa} and Quinn and 
Wald~\cite{QuWa}.  Practical procedures for regularizing the self-force in 
numerical calculations followed (e.g.,~\cite{BO_2000}).

The physical retarded field $p^{\text{ret}}_{\mu\nu}$ can be
conveniently split into regular ($R$) and singular ($S$) parts,
as first introduced by Detweiler and Whiting~\cite{DW_2003}.
The advantage of this split is that while the singular contribution
to the MP $p^{S}_{\mu\nu}$ satisfies the inhomogeneous field equations,
it does not contribute at all to the self-force.
On the other hand, the regular contribution, $p^{R}_{\mu\nu}$, is
a smooth, homogeneous solution to the field equations and,
through a projected gradient, is entirely 
responsible for the self-force.  Indeed, when interpreted this way,
the regular field can be thought of as an external field which sources
the deviation from geodesic motion on the background metric 
$g_{\mu \nu}$.  The motion of the particle is then geodesic on the spacetime
$g_{\mu \nu} + p^{R}_{\mu \nu}$.  The singular part of the MP is 
calculated analytically in 
Lorenz gauge, and an expansion provides the regularization 
parameters~\cite{BO_2000}.  The singular part is then subtracted from the 
full retarded field mode by mode in a spherical harmonic expansion, allowing 
the difference ($p_{\mu \nu}^{R}$) to converge.  While in 
principle~\cite{BO_2001,Barack_2009} the full retarded field could be 
calculated in a variety of gauges, in practice most 
calculations~\cite{BS_2007,BS_2009,BS_2010} have also used Lorenz gauge 
to find $p^{\text{ret}}_{\mu\nu}$.

We are developing a set of techniques and assembling a computer code to 
calculate with high accuracy the first-order MPs from a small compact 
object in a generic orbit about a Schwarzschild black hole.  The need for 
high accuracy is related to a set of arguments that have been made for years 
that EMRIs should be calculated through second-order in perturbation 
theory~\cite{Poisson_2002,Rosenthal_2006,Pound_2009,Galley_2011,Poisson_2011}.
One particular argument centers on calculating the phase evolution of an 
EMRI and using it in interpreting data from a detector.  For a small mass ratio 
$\epsilon = \mu/M$ (and in the absence of transient 
resonances~\cite{FH_2010} which may occur during EMRI evolution on a Kerr 
black hole), we expect that as an EMRI evolves through a detector passband 
the gravitational waveform will accumulate (schematically) a phase of
\be
\Phi = \kappa_1 \frac{1}{\epsilon} + \kappa_2 \epsilon^0 
+ \kappa_3 \epsilon^1 + \cdots ,
\ee
where the $\kappa$'s are coefficients of order unity that depend upon, among 
other things, the lower and upper limits in frequency of the detector 
response.  The first term reflects the dissipative effects of the first-order 
self-force in spurring a decay of the orbit.  The second term would result 
from second order in perturbation theory.  For example, with 
$\epsilon = 10^{-6}$, an EMRI might be observed to accumulate a total phase 
of $\Phi \sim 10^6$.  For matched filtering purposes we might need to compute 
the phase to an accuracy $\delta\Phi \lesssim 0.1$, and thus a fractional 
error of $\lesssim 10^{-7}$.  However, the error in phase in using the 
first-order calculation alone is $\sim \mathcal{O}(1)$.  Hence, the need for 
a second approximation to take full advantage of a detector output.  
\emph{But there is a corollary to this argument.}  We cannot possibly hope 
to make use of a second-order calculation if we have not already computed 
the first-order self-force to a relative accuracy \emph{much} better than 
$\mathcal{O}(\epsilon)$.  The requirement might be at least several orders 
of magnitude better than $\mathcal{O}(\epsilon)$ to make a second order 
calculation worthwhile (say $10^{-9}$ to $10^{-8}$ in the example).  
Furthermore, given that computation of the self-force is a numerically 
subtractive procedure, the first-order pre-regularization field contributions 
likely need to be known even more accurately (perhaps $10^{-11}$ to 
$10^{-10}$).

An accurate calculation strongly suggests use of Fourier decomposition and 
frequency domain (FD) methods, to gain the benefit of integrating ordinary 
differential equations.  Ultimately we are interested in the time-dependent 
self-force and must transfer back to the time domain (TD).  For that step, a 
lynchpin of the effort has been use of the recently developed method of 
extended homogeneous solutions (EHS)~\cite{BOS}, which allows partial Fourier 
series sums for the perturbations to avoid the Gibbs phenomenon and to 
converge exponentially even at the location of the point mass and despite 
loss of differentiability there.  This conclusion has guided others as 
well~\cite{akcay_2011,warburton_2012}.

Like other recent calculations, we want to determine the first-order 
MP in Lorenz gauge.  However, our approach is indirect.  
In an earlier paper~\cite{HE_2010}, we calculated 
radiative modes ($\ell\ge 2$) of the MP in Regge-Wheeler (RW) gauge by 
applying EHS to solutions of the master equations of the 
Regge-Wheeler-Zerilli (RWZ) formalism and then determining the metric from 
the master functions.  The new aspect of that work was being able to use FD 
techniques and nevertheless determine the metric amplitudes in the TD with 
accuracy right up to the location of the particle $r=r_p(t)$, an essential 
requirement for computing the self-force accurately.  EHS works by 
recognizing that the solution of a master equation with a moving singular 
source (in the RWZ case the source has both a delta function term and a 
derivative of delta function term) is a weak solution of the form 
$\Psi(t,r) = \Psi^{+}(t,r)\, \th \left[ r-r_p(t)\right] 
+ \Psi^{-}(t,r)\, \th \left[ r_p(t)-r \right]$, 
where $\Psi^{+}(t,r)$ and $\Psi^{-}(t,r)$ are differentiable solutions to the 
source-free master equation.  $\Psi^{+}$ and $\Psi^{-}$ are in turn obtained
as Fourier sums of properly-normalized Fourier-harmonic modes that solve 
the source-free master equation in the FD.  Since the functions in the 
separate Fourier sums are smooth everywhere, the lack of differentiability 
of $\Psi$ is captured entirely by the $\theta$ functions.  We can in turn then 
calculate the MPs from $\Psi$.  The result, however, is in RW gauge.  This 
paper, and a subsequent one, address the calculation of the infinitesimal 
gauge generator that transforms $p^{\text{RW}}_{\mu\nu}$ in RW gauge to its 
counterpart $p^{\text{L}}_{\mu\nu}$ in Lorenz gauge.  

This paper is restricted to finding the odd-parity part of the gauge 
generator, which for each spherical harmonic mode has an amplitude that is 
a solution to a single inhomogeneous wave equation.  While the wave equation 
is simple to express, what is more challenging is to find a way to 
generalize the underlying idea behind EHS to equations with 
non-compact source terms.  A substantial part of this paper is devoted to 
laying out two new analytic/numerical methods (\emph{method of partial 
annihilators} (PA) and \emph{method of extended particular solutions} (EPS)) 
we have developed for solving differential equations of this type.  
We present results from each of these methods, including a comparison of the 
two, showing that they agree to a high accuracy.
These techniques will play a central role in a subsequent paper where we 
present the more involved procedure, based on analysis by Sago, Nakano, 
and Sasaki~\cite{SNS_2003}, for determining the even-parity parts of 
the RW-to-Lorenz gauge generator. The Sago, Nakano, Sasaki approach takes
the gauge generator equations for even-parity, which are most naturally
expressed as a set of three, coupled equations for the vector spherical 
harmonic amplitudes, and transforms them to an altered set of equations in 
terms of different amplitudes.  The resulting equations are a hierarchical 
set of second-order equations, in which solutions to preceding steps in the 
hierarchy form source terms for subsequent steps. The result is a system 
of equations that, while containing more steps (amplitudes), lends itself to 
the immediate application of the techniques developed here.

One might ask, why two new methods?  In part, partial annihilators is the 
easier of the two methods to implement but requires that the partial 
annihilator operator be found.  In some applications that may be difficult.  
In contrast, EPS is straightforward if somewhat more involved in terms of the 
number of steps.  Ultimately, the main advantage of finding two methods is
that they provide a powerful check on each other and confirmation that the 
solution has been obtained.  We demonstrate this comparison in 
Sec.~\ref{sec:results}.

\begin{figure}
\vspace{-.5cm}
\includegraphics[scale=1]{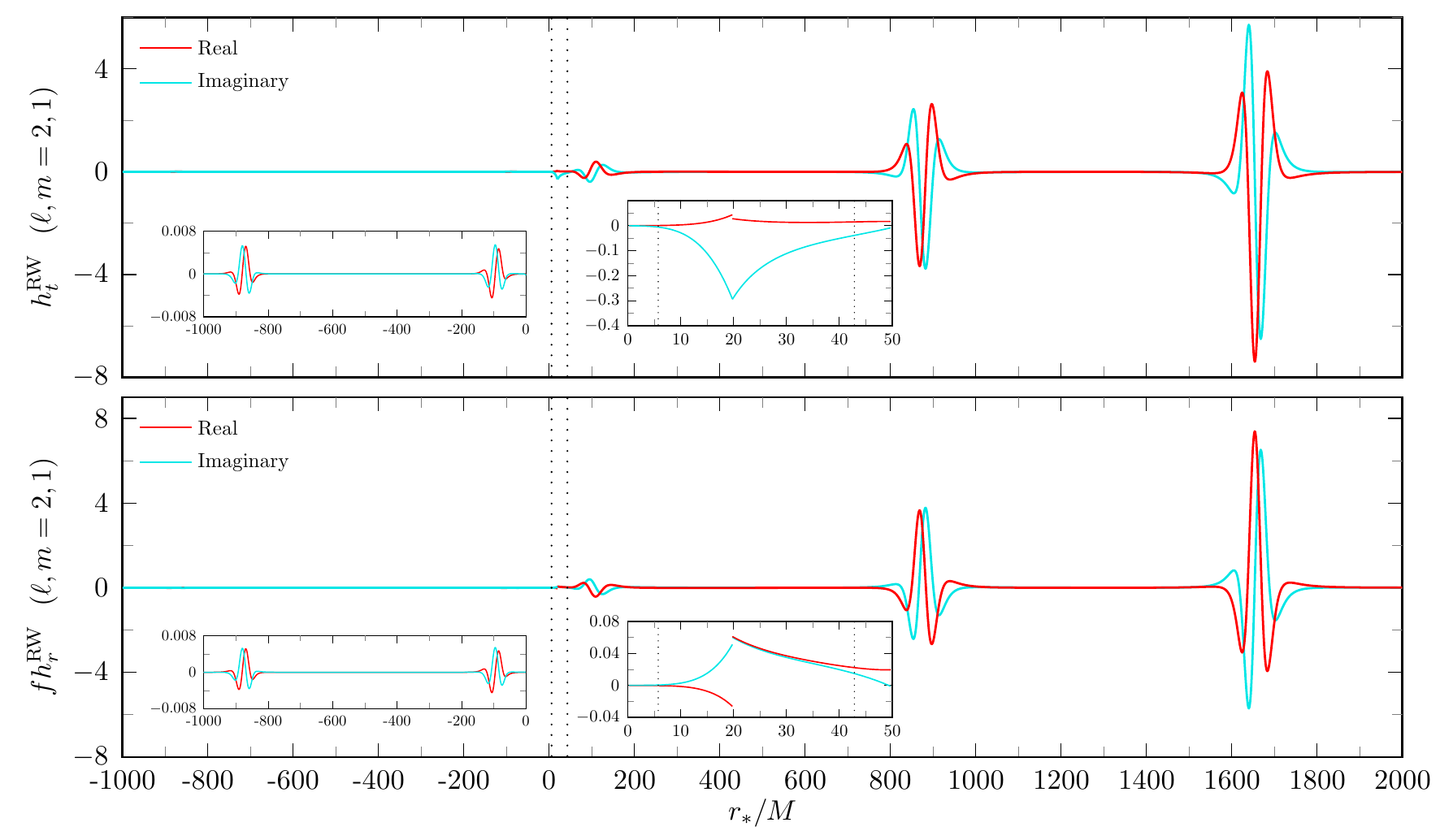}
\vspace{-.5cm}
\caption{
\label{fig:hthrRW}
The $\ell = 2$, $m = 1$ mode of the Regge-Wheeler gauge MP amplitudes 
$h_{t}^{\text{RW}}$ and $h_{r}^{\text{RW}}$.  The orbit is parameterized by 
eccentricity $e = 0.764124$ and semi-latus rectum $p = 8.75455$.  The plots 
show the real and imaginary parts of the MP amplitudes at time $t = 93.58$ 
(where $t = 0$ is at the periapsis).  Dotted vertical lines indicate limits 
of the source libration region.  Insets show the discontinuities at the 
location of the particle.  The lack of asymptotic flatness is evident as 
$r\rightarrow \infty$.  We plot $f h^{\text{RW}}_{r}$ so that the wave 
behavior near the horizon can be seen.}
\vspace{-.5cm}
\end{figure}

The discussion above begs the question, why the need to transform to Lorenz 
gauge?  In part we know that the MP amplitudes in Lorenz gauge are $C^0$ in 
the TD at the particle location.  As we discussed in Ref.~\cite{HE_2010} the 
MP amplitudes in RW gauge are one or two orders of continuity worse behaved 
(i.e., some amplitudes are $C^{-1}$, or discontinuous, and some have a 
delta function term at $r=r_p(t)$).  Fig.~\ref{fig:hthrRW} demonstrates this 
problem graphically for the $\ell = 2, m = 1$ (odd parity) RW amplitudes 
$h_{t}^{2,1}$ and $h_{r}^{2,1}$.  The insets show discontinuities in the MP 
amplitudes at the particle's location.  The plots also show the $\propto r$ 
growth in the wave pulse amplitude as $r \rightarrow \infty$, reflecting 
the fact that RW gauge is not asymptotically flat~\cite{TC_1967}.  
Transformation to Lorenz gauge not only improves the behavior of the modes 
at $r=r_p(t)$, it also removes the non-asymptotically-flat behavior seen in
the radiative modes of RW gauge.

Lastly, we note that for non-radiative ($\ell = 0,1$) modes, the RWZ formalism
breaks down.  Various researchers have used different gauges to solve the 
Einstein equations for these modes.  Considering generic motion on a 
Schwarzschild 
background, Zerilli \cite{Zerilli_1970} solved for these modes analytically
in his own gauge which exhibits $C^{-1}$ behavior for certain modes.  For
circular orbits Detweiler and Poisson \cite{DP_2003} showed how to transform 
Zerilli's solutions to Lorenz gauge.  Barack and Sago \cite{BS_2010} solved 
the Lorenz gauge field equations directly for these modes (and higher 
radiative modes).  The non-radiative modes provide a crucial contribution 
to the conservative piece of the self-force.  In this paper, though, we are 
solely concerned with transforming our RW gauge ($\ell \ge 2$) MP amplitudes 
to Lorenz gauge.  However, we note that while RW gauge is not defined 
for $\ell < 2$, there is no such restriction on the gauge transformation.  
Indeed, the work presented here can be directly extended to handle the 
transformations of non-radiative mode solutions from other gauges to Lorenz.

Throughout this paper we use the sign conventions and notation of 
Misner, Thorne, and Wheeler \cite{MTW_1973} and use units in which 
$c = G = 1$.  The background geometry is a non-rotating black hole which is
described in terms of Schwarzschild coordinates 
$x^{\mu} = (t,r,\theta, \varphi )$.

\section{Formalism}
\label{sec:formalism}

Consider the motion of a small compact object of mass $\mu$ in orbit about 
a static black hole of mass $M$, with $\mu/M \ll 1$.  The small body gives 
rise to a perturbation $p_{\mu\nu}$ in the metric relative to the 
Schwarzschild background,
\be
ds^2 = g_{\mu\nu} dx^{\mu} dx^{\nu} = 
-f(r) dt^2 + f(r)^{-1} dr^2 
+ r^2 \left( d\theta^2 + \sin^2\theta d\varphi^2 \right) ,
\ee
where $f(r) = 1 - 2M/r$.  We are concerned here only with the first-order 
part $p^{(1)}_{\mu\nu}$ of the MP in an expansion in powers of $\mu/M$ and 
accordingly simply set $p_{\mu\nu} = p^{(1)}_{\mu\nu}$.  The full 
non-stationary metric is then 
$\text{g}_{\mu\nu} = g_{\mu\nu} + p_{\mu\nu}(t,r,\theta,\varphi)$.  It often 
proves convenient to work with the trace-reversed MP
\be
\Bar{p}_{\mu\nu} = p_{\mu\nu} 
- \tfrac{1}{2} g_{\mu\nu} p_{\alpha\beta} g^{\alpha\beta} .
\ee
The Einstein field equations can be expanded about the background geometry.
Once the stress-energy tensor $T^{(0)}_{\a\b}$ associated with the motion of 
the small compact object has been specified,
the first-order linear equations
\be
\label{eq:linearEFE}
G^{(1)}_{\a\b}(\Bar{p}_{\mu\nu}) = 8 \pi T^{(0)}_{\a\b} ,
\ee
can be solved to determine $\Bar{p}_{\mu\nu}$.  The small body is 
approximated as a point particle and the stress tensor is that of the 
particle moving on a geodesic of the Schwarzschild black hole (i.e., 
zeroth-order approximation).  

In this work, we are interested in bound eccentric 
motion.  The orbit is parameterized by a pair of constants, which can 
alternately be taken to be Darwin's \cite{Darwin} eccentricity $e$ and 
dimensionless semi-latus rectum $p$, or the bounding values of radial 
motion $r_{\text{min}}$ and $r_{\text{max}}$, or the specific energy 
$\cal{E}$ and angular momentum $\cal{L}$.  The first integrals of motion 
are integrated using Darwin's curve parameter $\chi$ and are converted to 
functions of the form $r=r_p(t)$, $\varphi = \varphi_p(t)$, and 
$\tau = \tau_p(t)$ (with $\theta = \pi/2$).  The orbit 
has two fundamental frequencies with
a rate $\Omega_r$ associated with radial libration and a mean rate of 
azimuthal advance $\Omega_{\varphi}$.  See Darwin 
\cite{Darwin} for the orbital integration and Cutler et al.~\cite{CKP_1994} 
for first application to black hole perturbations from bound motion. 

\subsection{Harmonic decomposition of the field equations}

The usual route to solving the field equations (\ref{eq:linearEFE})
on a Schwarzschild background is to
introduce tensor spherical harmonics and decompose the equations into 
individual angular harmonic modes.  This approach was followed by Regge and 
Wheeler \cite{RW_1957} and Zerilli \cite{Zerilli_1970} in solving for 
certain master functions that represent the odd- and even-parity parts, 
respectively, of the gravitational field.  The MP is then
derived from these master functions in Regge-Wheeler gauge.  
In Ref.~\cite{HE_2010},
 we used a variant of this approach, with slightly different versions 
of the two master functions, and combined it with a new analytic/numerical 
method for finding convergent solutions near the location of the point 
particle.  

In that paper we followed Martel and Poisson \cite{MP_2005} in using their 
definitions of the tensor spherical harmonics.  We recap those definitions 
here.  The unit two-sphere $\mathcal{S}^2$ is covered by coordinates 
$(\theta, \varphi)$.  Upper-case Latin indices $A$, $B$, etc.~denote these 
two angular coordinates and associated tensor components.  
The coordinates $(t,r)$ cover the submanifold $\mathcal{M}^2$.  For these 
coordinates and tensor components we use lower-case Latin indices $a$, $b$, 
etc.  The full Schwarzschild spacetime is 
$\mathcal{M} = \mathcal{M}^2 \times \mathcal{S}^2$.  The usual scalar 
spherical harmonic functions are $Y^{\ell m}(\theta,\varphi)$.  From 
these we define even-parity $Y_A^{\ell m} = D_A Y^{\ell m}$ and
odd-parity $X_A^{\ell m} = - {\varepsilon_A}^B \, D_B Y^{\ell m}$ vector 
harmonics, where $D_A$ is the covariant derivative on $\mathcal{S}^2$.  The 
metric on the unit sphere is $\Omega_{AB} = \text{diag}[1, \sin^2\theta]$ and 
the Levi-Civita tensor is $\varepsilon_{AB}$.  Both are compatible with 
$D_A$: $D_C \, \Omega_{AB} = D_C \, \varepsilon_{AB} = 0$.  Martel and 
Poisson define the two even-parity tensor spherical harmonics as 
$\Omega_{AB} Y^{\ell m}$ and
$Y_{AB}^{\ell m}=\left[D_A D_B +\tfrac{1}{2}\ell (\ell +1)\right] Y^{\ell m}$,
where the latter is trace-free and thus differs from that used by Regge and 
Wheeler ($D_{A} D_{B} Y^{\ell m}$)~\cite{RW_1957}.  They take the odd-parity 
tensor harmonic to be
$X_{AB}^{\ell m} = - \tfrac{1}{2} 
\left( {\varepsilon_A}^C D_B + {\varepsilon_B}^C D_A \right) D_C Y^{\ell m}$,
which differs by a minus sign from that of~\cite{RW_1957}.

Using these definitions the MP is split into even- and 
odd-parity parts.  Each parity is decomposed into sums over $\ell$ and $m$ 
of their respective harmonics.  For even-parity there are seven amplitudes 
that Martel and Poisson define, which can be related to those of 
\cite{RW_1957} and \cite{Zerilli_1970}.  They are $h_{tt} = f H_{0}$, 
$h_{tr} = H_{1}$, $h_{rr} = H_{2} / f$, $j_{t} = h_{0}$, $j_{r} = h_{1}$, 
$G_{\rm here} = G_{\text{RW}}$, and $K_{\rm here} = K_{\text{RW}} 
- \ell (\ell + 1) G / 2$.  (Note: here and on many occasions later in the 
text we suppress the spherical harmonic scripts $\ell$ and $m$ for brevity 
when no confusion should arise.)  Then the even-parity MP is
\begin{align}
\label{eq:MPEven}
p_{ab} &= \sum_{\ell, m} h_{ab}^{\ell m} Y^{\ell m}, 
& p_{aB} &= \sum_{\ell, m} j_a^{\ell m} Y_B^{\ell m}, 
& p_{AB} &= r^2 \sum_{\ell, m}\Big( K^{\ell m} \O_{AB} Y^{\ell m} 
+  G^{\ell m} Y^{\ell m}_{AB} \Big).
\end{align}
For odd-parity there are three amplitudes, $ h_{t} = h_0 $, $ h_{r} = h_1 $, 
and $ h_{2}^{\rm here} = -h^{\text{RW}}_2 $, which are equivalent to the 
original definitions up to sign.  The odd-parity MP is then
\begin{align}
\label{eq:MPOdd}
p_{aB} = \sum_{\ell, m} h_a^{\ell m} X_B^{\ell m},  
\q \q
p_{AB} = \sum_{\ell, m} h_2^{\ell m} X^{\ell m}_{AB} ,
\end{align}
along with the fact that $p_{ab} = 0$.  For the balance of 
this paper we are only concerned with odd-parity.

Regge-Wheeler gauge places the (algebraic) condition on the metric that
$h_{2}^{\text{RW}} = 0$.  In this gauge the odd-parity field equations 
become
\begin{align}
\begin{split}
\label{eq:oddfieldeqns}
 - \pa_t \pa_r  h_r^{\text{RW}} + \pa_r^2  h_t^{\text{RW}} 
- \frac{2}{r} \pa_t  h_r^{\text{RW}} 
- \frac{\ell (\ell + 1) r - 4M}{r^3 f}  h_t^{\text{RW}} &= P^t, \\
\pa_t^2  h_r^{\text{RW}} - \pa_t \pa_r  h_t^{\text{RW}} 
+ \frac{2}{r} \pa_t  h_t^{\text{RW}} 
+ \frac{ (\ell + 2) (\ell - 1) f }{r^2}  h_r^{\text{RW}} &= P^r, \\
 -\frac{1}{f} \pa_t  h_t^{\text{RW}} + f \pa_r  h_r^{\text{RW}} 
+ \frac{2M}{r^2}  h_r^{\text{RW}} &= P ,
\end{split}
\end{align}
where the source amplitudes $P^t$, $P^r$, and $P$ (for each $\ell$ 
and $m$) are odd-parity projections of the stress tensor,
\begin{align}
P^a (t,r)
\equiv  \frac{16 \pi r^2}{\ell (\ell +1)}
\int T^{aB} X_{B}^{*} \, d\O ,
\q \q
P (t,r)  \equiv  
16 \pi r^4 \frac{(\ell - 2)!}{(\ell + 2)!} 
\int T^{AB} X^{*}_{AB}\, d\O .
\label{eq:sourceTerms}
\end{align}
We use an asterisk to denote complex conjugation.
The source amplitudes in turn satisfy the contracted Bianchi identity
\be
\frac{\partial P^t}{\partial t} + \frac{\partial P^r}{\partial r} 
+ \frac{2}{r} P^r - \frac{(\ell -1)(\ell +2)}{r^2} P = 0 ,
\ee
and the stress tensor itself is taken to be that of a particle in 
geodesic motion on the background geometry.

While in principle the coupled equations (\ref{eq:oddfieldeqns}) 
might be solved to yield the 
metric in RW gauge, the usual approach involves defining and using one of 
several odd-parity master functions and solving a lone wave equation (master 
equation) for this function.  The odd-parity part of the metric then is 
derived from the master function.  An equivalent master function 
representation is used for even-parity in RW gauge.  In our previous paper 
\cite{HE_2010} we used the odd-parity Cunningham-Price-Moncrief (CPM) 
function~\cite{CPM_1978},
 which we refer to here as $\Psi_{o}$.  This gauge-invariant master 
function is defined in Regge-Wheeler gauge by
\be
\label{eq:masterOdd}
\Psi_{o}(t,r) \equiv \frac{2 r}{(\ell -1)(\ell +2)}
\left[ \pa_r h_t^{\text{RW}} - \pa_t h_r^{\text{RW}} 
- \frac{2}{r} h_{t}^{\text{RW}} \right] .
\ee
In terms of the tortoise coordinate $r_{*} = r + 2 M \ln (r/2M -1)$, 
$\Psi_{o}$ satisfies the wave equation
\be
\label{eq:waveCPM}
\mathcal{W}_{2} \Psi_{o} (t,r) = S_{o} ,
\ee
where ${\cal W}_{2}$ is the spin-2 Regge-Wheeler operator,
a particular case of the spin-$s$ operator
\be
\label{eq:WsDef}
\mathcal{W}_{s} = -\frac{\pa^2}{\pa t^2}  + \frac{\pa^2}{\pa r_*^2}
- f \left[ \frac{\ell \l \ell + 1 \r}{r^2} + \frac{2 M (1 - s^2)}{r^3}\right] .
\ee
Later, we will have need for the Fourier transform 
of this operator ($\pa_{t} \to -i \o$), which
we will denote ${\cal L}_{s}$.
The source term for the master equation involves a combination of moments
of the stress tensor,
\be
S_{o} (t,r) \equiv \frac{ 2 r f}{(\ell -1)(\ell +2)}
\left[\frac{1}{f} \pa_t P^{r} + f \pa_r P^{t} + \frac{2M}{r^{2}} P^{t} \right] 
= \tilde G_{o} (t)\, \d \left[ r - r_{p}(t) \right] +
\tilde F_{o} (t)\, \d' \left[ r - r_{p}(t) \right] ,
\label{eq:SOdd}
\ee
and is a distribution (see~\cite{HE_2010} for details).
Once the CPM master function is known the MP amplitudes in 
RW gauge can be reconstructed via the expressions
\begin{align}
\label{eq:ReconstructOdd}
h_t^{\text{RW}} (t,r) = \frac{f}{2} \pa_r \l r \Psi_{o} \r 
- \frac{r^2 f}{(\ell -1)(\ell +2)} P^t,
\q \q
h_r^{\text{RW}} (t,r) = \frac{r}{2 f} \pa_t \Psi_{o} 
+ \frac{r^2}{(\ell -1)(\ell +2) f} P^r.
\end{align}
Their numerical determination in the time domain with a convergent 
and accurate behavior everywhere including the vicinity of the moving 
particle was the subject of our previous paper.

Because we have reason to consider it in what follows, it is worthwhile 
noting that the original master function of Regge and Wheeler, 
$\Psi_{\text{RW}}$, is not the CPM master function we use.  They are related 
by
\be
\label{eq:CPMRW}
\Psi_{\text{RW}} (t,r) = \frac{f}{r} h_r^{\text{RW}} (t,r) = 
\frac{1}{2} \pa_t \Psi_{o} + \frac{r}{(\ell -1)(\ell +2)} P^r .
\ee
The RW master function satisfies an almost identical wave equation,
\be
\label{eq:RWDE}
\mathcal{W}_{2} \Psi_{\text{RW}} (t,r) = S_{\text{RW}}  ,
\ee
with the only difference being the source term
\be
S_{\text{RW}} (t,r) \equiv \frac{f}{r}
\left[-P^{r} + f \pa_r P - \frac{2}{r} \l 1 - \frac{3M}{r} \r P \right] .
\label{eq:SRW}
\ee

\subsection{Gauge transformations}

The exact form of the field equations (\ref{eq:linearEFE}) will depend upon 
specifying a gauge.  
As mentioned in the Introduction, two frequent choices are Regge-Wheeler (RW)
gauge and Lorenz (L) gauge.  The small gauge generator $\Xi^{\mu}$ that 
transforms the coordinates 
$x^{\mu}_{\text{L}} = x^{\mu}_{\text{RW}} + \Xi^{\mu}$ between the two 
gauges is on the same order of magnitude as the MP, that is 
$|\Xi_{\mu}| \sim |\Bar{p}_{\mu \nu}| \ll 1$.  The MP then transforms as
\be
\label{eq:MPTrans}
\Bar{p}_{\mu \nu}^{\text{L}} = \Bar{p}_{\mu \nu}^{\text{RW}} 
- \Xi_{\mu | \nu} - \Xi_{\nu | \mu} + g_{\mu\nu} {\Xi^{\alpha}}_{| \alpha} ,
\ee
where stroke ${| \mu}$ indicates covariant differentiation with respect
to the background metric.  Lorenz gauge requires the following condition on 
the MP, 
\be
{{\Bar p}_{\mu \nu}^{\text{L}}}^{|\nu} = 0 .
\ee
Using this condition in Eq.~(\ref{eq:MPTrans}) then provides a wave 
equation that must be satisfied by the gauge generator,
\be
\label{eq:gaugeGen}
{\Xi_{\mu|\nu}}^{\nu} = {\Bar{p}_{\mu \nu}^{\text{RW}}}^{|\nu} .
\ee
A gauge generator that satisfies this
equation is unique only up to some $\Xi^{\prime}_{\mu}$ that satisfies 
the homogeneous version of (\ref{eq:gaugeGen}).  Specifying the initial data
and boundary values (if any) removes the residual gauge freedom and fully 
determines the gauge.

We consider next the spherical harmonic decomposition of the gauge vector. 
Momentarily considering again both even- and odd-parity, $\Xi_{\mu}$ can be
broken down into
\be
\Xi_{a} = \sum_{\ell,m} 
\left[ {\delta_a}^t \xi^{\ell m}_t (t,r) 
+ {\delta_a}^r \xi^{\ell m}_r (t,r) \right]
Y_{\ell m},
\q \q
\Xi_{A} = \sum_{\ell,m} \Big[ 
\xi^{\ell m}_{e} (t,r) Y_{A}^{\ell m} 
+ \xi^{\ell m}_{o} (t,r) X_{A}^{\ell m} \Big] .
\label{eq:xiExpand}
\ee
There are three even-parity amplitudes and one odd-parity amplitude.  We 
will concern ourselves with determining $\xi_t$, $\xi_r$, and $\xi_e$ in 
a subsequent paper.  In this paper we seek to obtain $\xi_o$.  Substituting 
the decomposition of $\Xi_{\mu}$ into Eq.~(\ref{eq:gaugeGen}), we find after 
a bit of calculation that $\xi_o$ satisfies the differential equation
\be
\mathcal{W}_1 \xi_{o} (t,r) = 2 f \Psi_{\text{RW}} + f P .
\label{eq:oddParityGauge}
\ee
Once the gauge generator amplitude is known,
we decompose Eq.~(\ref{eq:MPTrans}) in harmonic amplitudes
and see that the odd-parity MP 
amplitudes are transformed by
\begin{align}
\begin{split}
h_{t}^{\text{L}} (t,r) &= h_{t}^{\text{RW}} - \frac{\pa \xi_{o}}{\pa t}, \\
h_{r}^{\text{L}} (t,r) &= h_{r}^{\text{RW}} - \frac{\pa \xi_{o}}{\pa r} 
+ \frac{2}{r} \xi_{o},  \\
h_{2}^{\text{L}} (t,r) &= - 2 \xi_{o}.
\label{eq:push}
\end{split}
\end{align}

\subsection{Local nature of the metric perturbation
and gauge generator at $r=r_p(t)$}

The RHS of Eq.~(\ref{eq:oddParityGauge}) is singular at the location of the 
particle.  In this sense Eq.~(\ref{eq:oddParityGauge}) 
is very similar to Eq.~(\ref{eq:waveCPM}).
In Ref.~\cite{HE_2010} we examined Eq.~(\ref{eq:waveCPM})
to determine the local behavior of $\Psi_{o}$.
Assuming $\Psi_{o} = \Psi_{o}^{+}\, \th \left[ r - r_{p} (t) \right] +
\Psi_{o}^{-}\, \th \left[ r_{p} (t) - r \right]$, we calculated
jumps in the field, $\llbracket \Psi_{o} \rrbracket_{p}$, and in its
radial derivative, $\llbracket \pa_{r} \Psi_{o} \rrbracket_{p}$.
We use a subscript $p$ to indicate that a function of $r$ is evaluated
at the location of the particle, $r = r_{p}(t)$, becoming a function of time.

Following the same logic here, we postulate a form for 
the gauge amplitude of
$
\xi_{o}
=
\xi_{o}^{+}\, \th \left[ r - r_{p} (t) \right]
+
\xi_{o}^{-}\, \th \left[ r_{p} (t) - r \right].
$
Then, similar analysis to that found in Ref.~\cite{HE_2010}, indicates that
$\xi_{o}$ is $C^{0}$, i.e. $\llbracket \xi_{o} \rrbracket_{p} = 0$.
Further, we find the jump in the first radial derivative is
\be
\left\llbracket \pa_{r} \xi_{o} \right\rrbracket_{p} (t)
=
\frac{f_{p}}{f_{p}^{2} - {\dot r_{p}}^{2}}  p .
\ee
Here $p(t)$ comes from the source amplitude $P$.  All three source 
amplitudes are delta distributions with time dependent amplitudes: 
$P = p(t)\, \d [r - r_p(t)]$, $P^t = p^t(t)\, \d [r - r_p(t)]$, and
$P^r = p^r(t)\, \d [r - r_p(t)]$.

Having computed the expected jumps in $\xi_{o}$ and its radial derivative,
we can use Eq.~(\ref{eq:push}) to find the jumps in the Lorenz gauge MP 
amplitudes.  As with the fields $\Psi_{o}$ and $\xi_{o}$, we expect each 
MP amplitude to consist of left and right side differentiable functions 
that are joined at the location of the particle by Heaviside functions.
We calculated the jumps in $h_{t}^{\text{RW}}$ and $h_{r}^{\text{RW}}$ in 
Ref.~\cite{HE_2010}.  The discontinuities in the RW gauge MP amplitudes 
are exactly canceled out by terms arising from the derivatives of $\xi_{o}$ 
and all three Lorenz gauge amplitudes are $C^{0}$ as expected.  The jumps 
in their first derivatives are also analytically computable
(either by examining the jumps in the higher-order derivatives of 
$\xi_{o}$ or more simply by directly analyzing the Lorenz gauge
field equations).  We find
\begin{align}
\label{eq:LGJumps}
\left\llbracket \pa_{r} h_{t} \right\rrbracket_{p} (t)
= \frac{f_{p}^{2}}{f_{p}^{2} - {\dot r_{p}}^{2}}  p^{t},
\q \q
\left\llbracket \pa_{r} h_{r} \right\rrbracket_{p} (t)
= - \frac{1}{f_{p}^{2} - {\dot r_{p}}^{2}}  p^{r},
\q \q
\left\llbracket \pa_{r} h_{2} \right\rrbracket_{p} (t)
= - \frac{2 f_{p}}{f_{p}^{2} - {\dot r_{p}}^{2}}  p .
\end{align}
We use these expressions later (see Fig.~\ref{fig:convVsN}) 
as a powerful check that we have correctly
solved the gauge transformation equations to high accuracy.

\section{Two EHS-like methods for equations with non-compact sources}
\label{sec:twomethods}

In Ref.~\cite{HE_2010} we solved Eq.~(\ref{eq:waveCPM}) for a variety of 
eccentric orbits.  We used a FD approach to find the Fourier harmonic 
modes of $\Psi_{o}$ and transformed back to the TD using the EHS method.  
The EHS method was first applied to wave equations with delta function 
sources.  It allows TD reconstruction of the spherical harmonic amplitudes 
with exponential convergence, circumventing the Gibbs phenomenon that 
otherwise arises from solving equations with discontinuous or singular 
sources.  Our application of the method also demonstrated it could be 
applied to sources with a derivative of a delta function.

With only a change in spin parameter, Eq.~(\ref{eq:oddParityGauge}) 
has a similar differential operator as Eq.~(\ref{eq:waveCPM}).  Where the 
two equations differ markedly is in their source terms.  While 
the source in Eq.~(\ref{eq:waveCPM}) is point-singular and compact, the
source in Eq.~(\ref{eq:oddParityGauge}) is both distributional and 
non-compact.  We can use the linearity of the equation to split off the 
singular part and split the generator into two parts,
$\xi_{o} = \xi_{o}^{\text{ext}} + \xi_{o}^{\text{sing}}$,
that satisfy separate equations,
\begin{align}
{\cal W}_{1} \xi_{o}^{\text{sing}} (t,r)
&=
 f_{p} p(t)\, \d \left[ r - r_{p} (t) \right], 
\label{eq:xiSing}
 \\
{\cal W}_{1} \xi_{o}^{\text{ext}} (t,r)
&=
2 f \Psi_{\text{RW}} .
\label{eq:xiExtended}
\end{align}
While the former equation can be solved using the EHS method, the latter's 
extended source term is more problematic.  The extended source is both 
non-compact and has a time-dependent discontinuity that moves periodically 
between $r_{\text{min}}$ and $r_{\text{max}}$ as the particle orbits.  In 
this section we present two equivalent methods for solving 
Eq.~(\ref{eq:xiExtended}) using FD methods, both of which provide 
exponential convergence upon returning to the TD.

As discussed earlier, an eccentric orbit on Schwarzschild provides two 
fundamental frequencies.  When we Fourier transform 
Eq.~(\ref{eq:xiExtended}), we have a two-fold countably infinite frequency 
spectrum,
\be
\o \equiv \o_{mn} = m \O_{\varphi} + n \O_{r}, \q \q m,n \in \mathbb{Z}.
\ee
The Fourier transform and standard TD reconstruction of 
$\xi_{o}^{\text{ext}} (t,r)$ is then
\be
\tilde \xi_{o}^{\text{ext}} (r) \equiv \frac{1}{T_r} \int_0^{T_r} dt \ \xi_{o}^{\text{ext}} (t,r)
\, e^{i \o t},
\q \q
\xi^{\text{ext}}_{o} (t,r) = \sum_{n=-\infty}^{\infty} 
\tilde \xi^{\text{ext}}_{o}(r) \, e^{-i \o t}.
\ee
Note that in addition to the already suppressed indices $\ell$ and $m$,
FD quantities have a third implied index, $n$.
Equivalent expressions are used for the Fourier transforms and 
series representations of the other fields that we consider below.
Note that while we use the standard tilde \~ \ notation with
$\tilde \xi^{\text{ext}}_{o}$ to indicate a FD quantity, 
for other quantities we try to maintain consistency with 
previous literature by changing the base 
symbol.
For the TD function $\Psi_{\text{RW}}$, we
write $R_{\text{RW}}$ in the FD. Similarly, for its  TD 
source term $S_{\text{RW}}$, we write  $Z_{\text{RW}}$ in the FD.

\subsection{First approach: partial annihilator method}
\label{sec:partialannihilator}

Our first method for solving Eq.~(\ref{eq:xiExtended}) is a generalization
of the standard method of annihilators used for solving inhomogeneous 
differential equations.  It hinges on finding an ``annihilator,'' a
differential operator which gives a vanishing result after acting
on the source.  Then, one can act with the annihilator on both sides of the 
differential equation.  What results is a homogeneous differential equation
of higher order.  Our strategy for solving Eq.~(\ref{eq:xiExtended}) is 
essentially the same, except that because the initial source is discontinuous 
the operator that we find does not completely annihilate the RHS but instead 
converts it to a distribution.  Hence, we refer to the operator as 
a \emph{partial annihilator}.

The RHS of Eq.~(\ref{eq:xiExtended}) is well suited to the method
of partial annihilators because the Regge-Wheeler variable satisfies its own 
wave equation with a point-singular source, (\ref{eq:RWDE}).  Therefore, 
upon dividing Eq.~(\ref{eq:xiExtended}) by $f$, we can take ${\cal W}_{2}$ as 
the partial annihilator and act on both sides of the equation
\be
f {\cal W}_{2} \left( \frac{1}{f}\, {\cal W}_{1} \xi (t,r) \right)
 = 2 f S_{\text{RW}} (t, r_{p}(t)) 
 = 2 f
  \l \tilde G_{\text{RW}} (t)\, \d \left[ r - r_{p}(t) \right]
 +
 \tilde F_{\text{RW}} (t)\, \d' \left[ r - r_{p}(t) \right] \r .
\label{eq:PATD}
\ee
For simplicity here and in the remainder of this section we drop the 
$^{\text{ext}}_{o}$ tags.  We have multiplied back through by $f$ to 
ensure that the leading-order derivatives have unit coefficients.
This differential equation is now fourth-order, but its source is 
point-singular.  This allow us to solve it using the EHS method, generalized
to fourth-order equations.  The specific form of the source in 
Eq.~(\ref{eq:PATD}) is given by Martel \cite{Martel_2004}, though we 
assume that both $\tilde G_{\text{RW}}$ and $\tilde F_{\text{RW}}$ have
been evaluated at $r = r_{p}(t)$.

We Fourier transform Eq.~(\ref{eq:PATD}) to obtain the FD equation 
\be
f {\cal L}_{2} \left( \frac{1}{f}\, {\cal L}_{1} \tilde \xi (r) \right)
 = 2 f Z_{\text{RW}} (r).
\label{eq:PAFD}
\ee
The Fourier transform averages the point source motion in time and produces
$Z_{\text{RW}}(r)$ which has support only within the source libration region 
$r_{\rm min}<r<r_{\rm max}$.

There are four linearly independent homogeneous solutions to
Eq.~(\ref{eq:PAFD}).   Two of these are the solutions to the 
second-order equation ${\cal L}_{1} \tilde \xi = 0$ and we denote them by 
$\tilde{\xi}^{\pm}_{h2}$.  One behaves asymptotically as an outgoing wave 
at infinity while the other is downgoing at the horizon
\be
\tilde \xi_{h2}^{-} 
\sim e^{-i \o r_{*}} \q (r \to 2M), \q \q
\tilde \xi_{h2}^{+} 
\sim e^{i \o r_{*}}  \q (r \to \infty).
\ee
The other two solutions only satisfy the full fourth-order 
equation ${\cal L}_{2} (f^{-1} \, {\cal L}_{1} \tilde \xi ) = 0$.
As such we give them the label ${h4}$ and asymptotic analysis shows that
\be
\tilde \xi_{h4}^{-} 
\sim f(r)\, e^{-i \o r_{*}}  \q (r \to 2M),  \q \q
\tilde \xi_{h4}^{+}
 \sim r\, e^{i \o r_{*}}  \q (r \to \infty) .
\label{eq:homog4th}
\ee
These four solutions form a basis spanning the space of homogeneous
solutions of Eq.~(\ref{eq:PAFD}).  The particular solution will be a 
linear combination of these with variable coefficients
\be
\tilde \xi_{p} (r)
 = c_{h2}^{-} (r)\, 
 \tilde \xi_{h2}^{-} (r) 
+ c_{h2}^{+} (r)\, 
\tilde  \xi_{h2}^{+} (r)
+ c_{h4}^{-} (r)\, 
\tilde  \xi_{h4}^{-} (r) 
+ c_{h4}^{+} (r)\, 
\tilde  \xi_{h4}^{+} (r).
\ee
The four normalization functions $c_{h2/h4}^{\pm} (r)$ are fixed by the 
method of variation of parameters, which entails solving the equations
\be
\label{eq:varOfPar}
\frac{d c^{\pm}_{h2/h4}}{d r_{*}} 
= 2 f Z_{\text{RW}} (r) \frac{W^{\pm}_{h2/h4} (r)}{W(r)}.
\ee
Here $W(r)$ is the Wronskian and $W^{\pm}_{h2/h4} (r)$ is the ``modified 
Wronskian'' (Cramer's rule), which is the Wronskian with the column 
corresponding to the $\xi_{h2/h4}^{\pm} (r)$ homogeneous solution replaced 
by the column vector $(0, 0, 0, 1)$.  Note that because the differential 
operator in Eq.~(\ref{eq:PATD})
is written in terms of $r_{*}$, the derivatives within the Wronskian must
also be taken with respect to $r_{*}$ and the LHS of Eq.~(\ref{eq:varOfPar})
is a derivative taken with respect to $r_{*}$.
For the two ``$+$'' equations, the integral form of Eq.~(\ref{eq:varOfPar}) is 
(we change the variable of integration to $r$ and see the factor of $f$
cancel)
\be
c^{+}_{h2/h4} (r) 
= 2 \int_{r_{\rm min}}^{r}
\bigg[ 
\frac{1}{T_{r}}
\int_{0}^{T_{r}}  \Big( 
\tilde G_{\text{RW}} (t)\, \d \left[ r' - r_{p}(t) \right] 
+\tilde F_{\text{RW}} (t)\, \d' \left[ r' - r_{p}(t) \right]
 \Big) e^{i \o t} dt \bigg] 
 \frac{W^{+}_{h2/h4} (r')}{W(r')} dr'.
\ee
Likewise, for the two ``$-$'' equations (note the change on the limits of 
integration),
\be
c^{-}_{h2/h4} (r) 
= 2 \int_{r}^{r_{\rm max}}
\bigg[ 
\frac{1}{T_{r}}
\int_{0}^{T_{r}}  \Big( 
\tilde G_{\text{RW}} (t)\, \d \left[ r' - r_{p}(t) \right] 
+\tilde F_{\text{RW}} (t)\, \d' \left[ r' - r_{p}(t) \right]
 \Big) e^{i \o t} dt \bigg] 
 \frac{W^{-}_{h2/h4} (r')}{W(r')} dr'.
\ee
The EHS method requires knowing only the terminal values of the four 
functions $c^{\pm}_{h2/h4}(r)$, i.e.,
$C^{+}_{h2/h4} = c^{+}_{h2/h4}(r_{\text{max}})$ and 
$C^{-}_{h2/h4} = c^{-}_{h2/h4}(r_{\text{min}})$.  Switching the order of 
integration and integrating by parts, we find
\be
C^{\pm}_{h2/h4} = 
\frac{2}{T_{r}}
\int_{0}^{T_{r}}  
\Bigg\{ 
\tilde G_{\text{RW}} (t)  
\frac{W^{\pm}_{h2/h4} (r_{p})}{W(r_{p})} 
- \tilde F_{\text{RW}} (t) 
\left[ 
-  \frac{W^{\pm}_{h2/h4} (r_{p} )}{W(r_{p})^{2}} \pa_{r} W (r_{p})
+ \frac{ \pa_{r} W^{\pm}_{h2/h4} (r_{p})}{W (r_{p})} \right]
\Bigg\} e^{i \o t} dt.
\ee
At this point we define the EHS in the FD to be 
\be
\tilde  \xi_{h}^{-} (r) \equiv C^{-}_{h2}\tilde  \xi_{h2}^{-} (r) 
+ C^{-}_{h4} \tilde \xi_{h4}^{-} (r), 
\q \q
\tilde \xi_{h}^{+} (r) \equiv C^{+}_{h2} \tilde \xi_{h2}^{+} (r) 
+ C^{+}_{h4} \tilde \xi_{h4}^{+} (r), 
\ee
and the EHS in the TD are defined by the Fourier
sums (recall the suppressed $\ell, m, n$ indices)
\be
\xi^{\pm} (t,r) \equiv \sum_{n} \tilde  \xi_{h}^{\pm} (r) e^{-i \o t}.
\ee
The extension of these solutions to $r=r_{p}(t)$ then gives the desired
solution to Eq.~(\ref{eq:xiExtended}),
\be
\xi_{o}^{\text{ext}} (t,r) =
\xi^{+} (t,r)\, \th \left[ r - r_{p} (t) \right]
+
\xi^{-} (t,r)\, \th \left[ r_{p} (t) - r \right].
\label{eq:xiPartAnn}
\ee

\subsection{Second approach: method of extended particular solutions}
\label{sec:epsmethod}

Now we look for a solution to Eq.~(\ref{eq:xiExtended}) 
that does not require a partial annihilator.
In the FD the equation transforms to
\be
{\cal L}_{1} \tilde \xi_{o}^{\text{ext}} (r)
= 2 f R_{\text{RW}}.
\label{eq:xiFD}
\ee
Again, for notational simplicity we drop the $^{\rm ext}_{o}$ tags
for the remainder of this section.  In the end we want solutions to
Eq.~(\ref{eq:xiFD}) that allow us to form an exponentially converging 
solution to Eq.~(\ref{eq:xiExtended}) when we transfer to the TD.  This 
will require a new technique which we call
\emph{extended particular solutions}, and is closely analogous to the
EHS method.  First, though, we 
consider how to get the correct causal solution to Eq.~(\ref{eq:xiFD})
from a ``standard'' approach.

In the subsequent sections we make a distinction between quantities with
$^{\infty}$ and $^{H}$ tags which designate functions computed
from a ``standard'' RHS source (Eq.~(\ref{eq:RStdRW}) below) 
and those quantities with
$^{+}$ and $^{-}$ tags which designate functions computed 
from an ``extended'' RHS source (Eq.~(\ref{eq:R_EHS}) below). 
Because the homogeneous solutions do not depend on the source, we always
tag them with $^{+}$ or $^{-}$.  We distinguish between
particular and homogeneous solutions by using the respective subscripts
$_{p}$  and $_{h}$.

\vspace{-.2cm}
\subsubsection{Finding standard FD solutions with causal boundary conditions}

By examining the source, $2f R_{\text{RW}}$, and the differential operator,
${\cal L}_{1}$, we can obtain asymptotic and Taylor expansions of the 
particular solution $\tilde \xi_{p}$ near infinity and the horizon, 
respectively.  The expansions are useful numerically but for our purposes 
here we need only consider the leading asymptotic dependence.
(See App.~\ref{sec:asympExp} for discussion of the asymptotic expansion 
($r \rightarrow \infty$) of $\tilde \xi_{p}$ and how it couples to the 
expansion of $R_{\text{RW}}$.)

Consider first the spatial infinity side.  Let the RW function have an 
asymptotic amplitude $C^{+}_{\text{RW}}$, so 
$R_{\text{RW}} = C^{+}_{\text{RW}} e^{ i \o r_{*}}$ as $ r_{*} \to  \infty$.
We then make the ansatz that 
$\tilde \xi_{p} = C^{\infty}_{p} r e^{i \o r_{*}}$ as $r_{*} \to \infty$.  
Using an asymptotic approximation to Eq.~(\ref{eq:xiFD}) we find
\begin{align}
\l \frac{d^{2}}{d r_{*}^{2}} + \o^{2} \r 
\l C^{\infty}_{p} r e^{ i \o r_{*}} \r
= 2 C^{+}_{\text{RW}}  e^{ i \o r_{*}}  \q \q
\Rightarrow
\q\q
C^{\infty}_{p} 
= \frac{1}{i \o} C^{+}_{\text{RW}}.
\label{eq:xiCPlus}
\end{align}
Therefore, the asymptotic form of $\tilde \xi_{p}^{\infty}$ is 
\be
\tilde \xi_{p}^{\infty} 
= - \frac{i}{\o} C^{+}_{\text{RW}} r  e^{i \o r_{*}}, \q \q 
r \to  \infty.
\label{eq:xipInfty}
\ee
Next we consider the horizon side.  The RW function is asymptotically
$R_{\text{RW}} = C^{-}_{\text{RW}} e^{-i\o r_{*}}$ as $r_{*} \to -\infty$.
In this case we expect the particular solution to behave as
$\tilde \xi_{p} = C^{H}_{p}\, f e^{-i \o r_{*}}$ as $r_{*} \to -\infty$. 
Again, acting with the near-horizon leading parts of the differential 
operator we find
\begin{align}
\l \frac{d^{2}}{d r_{*}^{2}} + \o^{2} \r 
\l C^{H}_{p} f e^{ -i \o r_{*}} \r
= 2 f C^{-}_{\text{RW}} e^{ -i \o r_{*}}  \q 
\Rightarrow
\q
C^{H}_{p} 
=  2 \l \frac{1}{4M^{2}}- \frac{i \o}{M} \r^{-1} C^{-}_{\text{RW}}.
\label{eq:xiCMinus}
\end{align}
Therefore, the near-horizon form of $\tilde \xi_{p}^{H}$ is
\be
\tilde \xi_{p}^{H} 
= 
2 \l \frac{1}{4M^{2}}- \frac{i \o}{M} \r^{-1} C^{-}_{\text{RW}}
 f  e^{-i \o r_{*}}, \q \q 
r \to 2M.
\label{eq:xipH}
\ee

We can use Eqs.~(\ref{eq:xipInfty}) and (\ref{eq:xipH}) to set boundary 
conditions (B.C.'s) for two separate integrations of the inhomogeneous 
differential equation, (\ref{eq:xiFD}) (yielding two different particular 
solutions that differ by some homogeneous solution).  This integration 
requires the source $R_{\text{RW}}$, which is itself the solution to the 
differential equation
\be
{\cal L}_{2} R_{\text{RW}} (r)  = Z_{\text{RW}} (r).
\label{eq:RWeq}
\ee
We find it by variation of parameters, which yields
\be
R^{\text{std}}_{\text{RW}} (r) 
= c^{+}_{\text{RW}} (r) \hat R^{+} (r) + c^{-}_{\text{RW}} (r) \hat R^{-} (r),
\label{eq:RStdRW}
\ee
where $\hat R^{\pm} (r)$ are unit-normalized homogeneous solutions to 
Eq.~(\ref{eq:RWeq}).  Note that 
$c_{\text{RW}}^{+} (r \ge r_{\text{max}}) = C^{+}_{\text{RW}}$
and 
$c_{\text{RW}}^{-} (r \le r_{\text{min}}) = C^{-}_{\text{RW}}$.  Furthermore, 
the solution $R^{\text{std}}_{\text{RW}} (r)$ is \emph{not} the same as the 
EHS to Eq.~(\ref{eq:RWeq}).

Having solved Eq.~(\ref{eq:RWeq}) for the source term in 
Eq.~(\ref{eq:RStdRW}), and determined the 
B.C.'s, we are ready to solve Eq.~(\ref{eq:xiFD}).  The idea is to find the 
two different particular solutions, neither of which has the proper causal 
behavior, and then correct for the acausality by adding appropriate
homogeneous solutions.  The result of this process is a 
solution to Eq.~(\ref{eq:xiFD}) with causal behavior on both sides.  The 
details follow in a series of steps. 

\begin{enumerate}
\item 
{\sc Solve for the particular solution from the spatial-infinity side
(see Fig.~\ref{fig:xiPInfty}).}

We set a B.C. to Eq.~(\ref{eq:xiFD}) using 
Eq.~(\ref{eq:xipInfty}) and integrate  to large negative
$r_{*}$ using Eq.~(\ref{eq:RStdRW}) as the source.
Although the starting B.C.
specified no homogeneous contribution, homogeneous
solutions on the horizon side will be excited.
See the left side of Fig.~\ref{fig:xiPInfty}. 
The particular solution integrated from 
the spatial infinity side has the asymptotic behavior
\be
\tilde \xi_{p}^{\infty} 
=
\left\{
\begin{array}{ll}
C^{\infty}_{p} r e^{i \o r_{*}}, 
& r_{*} \to +\infty, \\
C^{H}_{p} f e^{-i \o r_{*}} 
+ \k^{+} e^{i \o r_{*}} + \k^{-} e^{-i \o r_{*}},
&
 r_{*} \to -\infty .
\end{array}
\right.
\label{eq:xiPInftyOnH}
\ee  
The term with coefficient $C^{H}_{p}$ is the part directly dependent on the 
source and it is sub-dominant in comparison to the homogeneous solutions.
The coefficients $\k^{\pm}$ are to-be-determined.
Importantly, $\k^{+} e^{i \o r_{*}}$ is an acausal term (upgoing from the 
past horizon).

\begin{figure}[h!]
\includegraphics[scale=.95]{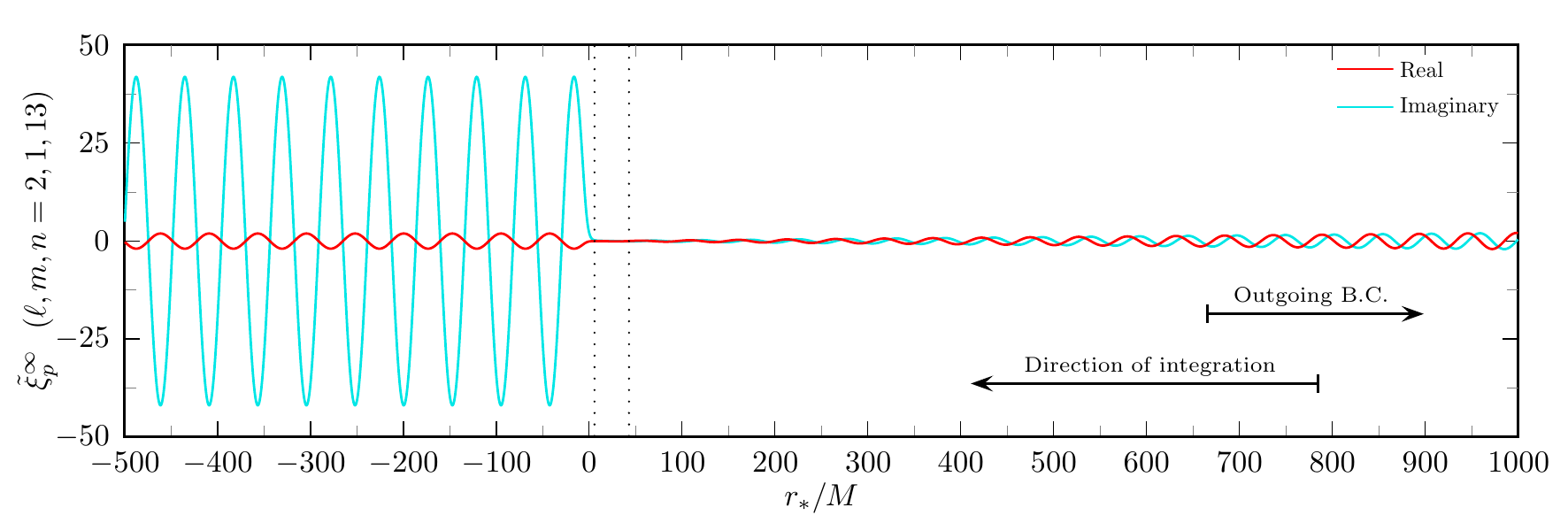}
\caption{
\label{fig:xiPInfty}
Integration from large $r_{*}$ of the particular solution, 
$\tilde \xi_{p}^{\infty}$.
Dotted lines indicate the source libration region.
}
\vspace{-.3cm}
\end{figure}

\item {\sc Solve for the particular solution from the horizon side
(see Fig.~\ref{fig:xiPH}).}

We set a B.C. to Eq.~(\ref{eq:xiFD}) using Eq.~(\ref{eq:xipH})
and integrate to large positive
$r_{*}$ using Eq.~(\ref{eq:RStdRW}) as the source.
 Although the starting B.C.
specified no homogeneous contribution, homogeneous
solutions on the spatial infinity side will be excited.
The effect can be seen on the right side of Fig.~\ref{fig:xiPH}.  
The particular solution dominates, but the constant offset between real
and imaginary parts indicates the presence of 
asymptotically-constant-amplitude homogeneous terms.  Analysis shows that 
the particular solution integrated from the horizon will behave as
\be
\tilde \xi_{p}^{H} 
=
\left\{
\begin{array}{ll}
C^{H}_{p} f e^{ -i \o r_{*}},
& r_{*} \to -\infty, \\
C^{\infty}_{p} r e^{i \o r_{*}} 
+ \la^{-} e^{-i \o r_{*}} + \la^{+} e^{i \o r_{*}},
&
 r_{*} \to +\infty.
\end{array}
\right.
\label{eq:xiPHOnInfty}
\ee  
The coefficients $\la^{\pm}$ are to-be-determined and again we find an 
acausal term ($\la^{-} e^{-i \o r_{*}}$), which in this case is ingoing 
from past null infinity.

\begin{figure}[h!]
\includegraphics[scale=.95]{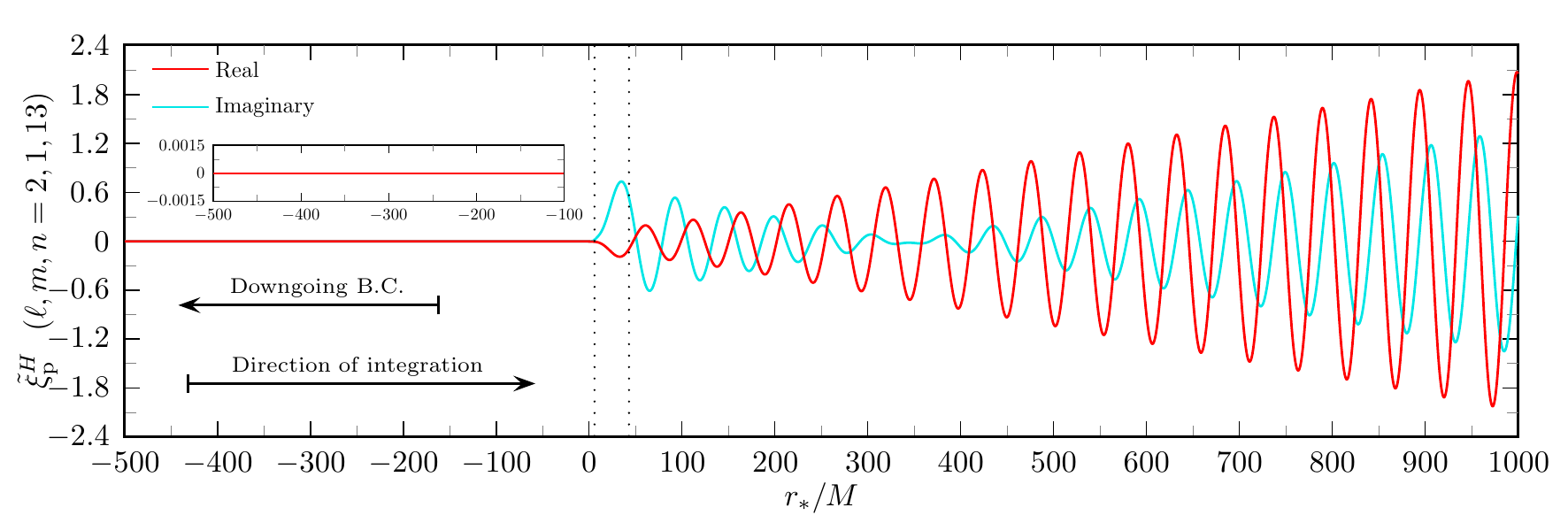}
\caption{
\label{fig:xiPH}
Integration from large negative
$r_{*}$ of the particular solution, $\tilde \xi_{p}^{H}$.
Dotted lines indicate the source libration region.
}
\vspace{-.3cm}
\end{figure}

\item {\sc Solve for the homogeneous solution from the spatial infinity
side.}

Next, we set an outgoing B.C. to the homogeneous version of 
Eq.~(\ref{eq:xiFD}) at large positive $r_{*}$.
We integrate to solve the scattering problem for
reflection and transmission amplitudes $R^{+}$ and $T^{+}$
\cite{Chandra_1983}.  In the terminology of Gal'tsov \cite{Galtsov}
this is an ``up'' mode,
\be
\tilde \xi_{h}^{+} 
=
\left\{
\begin{array}{ll}
T^{+} e^{i \o r_{*}},
& r_{*} \to +\infty, \\
R^{+} e^{-i \o r_{*}} +  e^{i \o r_{*}},
&
 r_{*} \to -\infty.
\end{array}
\right.
\label{eq:xiHPlus}
\ee  
Scaled appropriately, this solution can be added to
Eq.~(\ref{eq:xiPInftyOnH}) to remove its acausality.

\item {\sc Solve for the homogeneous solution from the horizon side.}

We set a downgoing B.C. to the homogeneous version of 
Eq.~(\ref{eq:xiFD}) at large negative $r_{*}$ and
integrate to solve the scattering problem for reflection and transmission 
amplitudes $R^{-}$ and $T^{-}$.  This is an ``in'' mode,
\be
\tilde \xi_{h}^{-} 
=
\left\{
\begin{array}{ll}
 T^{-} e^{-i \o r_{*}},
& r_{*} \to -\infty, \\
R^{-} e^{i \o r_{*}} +  e^{-i \o r_{*}},
&
 r_{*} \to +\infty.
\end{array}
\right.
\label{eq:xiHMinus}
\ee  
Scaled appropriately, this solution can be added to
Eq.~(\ref{eq:xiPHOnInfty}) to remove its acausality.

\item {\sc Resolve the acausality in the particular solutions 
(see Fig.~\ref{fig:xi_p_Std}).}
\label{resolveStep}

The acausal piece in Eq.~(\ref{eq:xiPInftyOnH}) is
$\k^{+} e^{i \o r_{*}}$.
Using Eq.~(\ref{eq:xiHPlus}), we can remove this by subtracting 
$\k^{+} \tilde \xi^{+}_{h}$,
\be
\tilde \xi^{\infty}_{p} - \k^{+} \tilde \xi^{+}_{h}
=
\left\{
\begin{array}{ll}
C^{\infty}_{p} r e^{i \o r_{*}} 
- \k^{+} T^{+}  e^{i \o r_{*}},
& r_{*} \to +\infty, \\
C^{H}_{p} f e^{-i \o r_{*}} 
+ \k^{-} e^{-i \o r_{*}}
- \k^{+} R^{+} e^{-i \o r_{*}},
&
 r_{*} \to -\infty.
\end{array}
\label{eq:xiPA}
\right.
\ee
The acausal piece in Eq.~(\ref{eq:xiPHOnInfty}) is
$\la^{-} e^{-i \o r_{*}}$.
Using Eq.~(\ref{eq:xiHMinus}), we can remove this by subtracting 
$\la^{-} \tilde \xi^{-}_{h}$,
\be
\tilde \xi^{H}_{p} - \la^{-} \tilde \xi^{-}_{h}
=
\left\{
\begin{array}{ll}
C^{H}_{p} f e^{ -i \o r_{*}}
- \la^{-} T^{-} e^{ -i \o r_{*}},
& r_{*} \to -\infty, \\
C^{\infty}_{p} r e^{i \o r_{*}} 
 + \la^{+} e^{i \o r_{*}}
- \la^{-} R^{-} e^{ i \o r_{*}} ,
&
 r_{*} \to +\infty.
\end{array}
\right.
\label{eq:xiPB}
\ee  
Eqs.~(\ref{eq:xiPA}) and (\ref{eq:xiPB}) are both solutions
to Eq.~(\ref{eq:xiFD}) and both satisfy the 
causal nature of the problem.  Therefore they must be equal.
In order to form them, we must know $\k^{+}$ and $\la^{-}$.
We find them by setting Eqs.~(\ref{eq:xiPA}) and (\ref{eq:xiPB}) 
and their first derivatives equal at any point,  
\begin{align}
\tilde \xi^{H}_{p} - \la^{-} \tilde \xi^{-}_{h} &=
\tilde \xi^{\infty}_{p} - \k^{+} \tilde \xi^{+}_{h},
\label{eq:XiEq}
 \\
\pa_{r_{*}} \tilde \xi^{H}_{p} - \la^{-} \pa_{r_{*}} \tilde \xi^{-}_{h} &=
\pa_{r_{*}} \tilde \xi^{\infty}_{p} - \k^{+} \pa_{r_{*}} \tilde \xi^{+}_{h} .
\label{eq:diffXiEq}
\end{align}
We solve these equations for $\k^{+}$ and $\la^{-}$ and 
form  $\tilde \xi^{H}_{p} - \la^{-} \tilde \xi^{-}_{h}$ and 
$\tilde \xi^{\infty}_{p} - \k^{+} \tilde \xi^{+}_{h}$, which are equivalent.
In principle one could pick any point and expect the same result.  In 
practice, slight numerical differences occur.  In fact, we use several points 
to determine these constants and use the discrepancies that are found as a 
measure of the order of magnitude of the error.  Cumulative numerical error 
in the solutions is addressed in Sec.~\ref{sec:results}.

\begin{figure}[h!]
\includegraphics[scale=1]{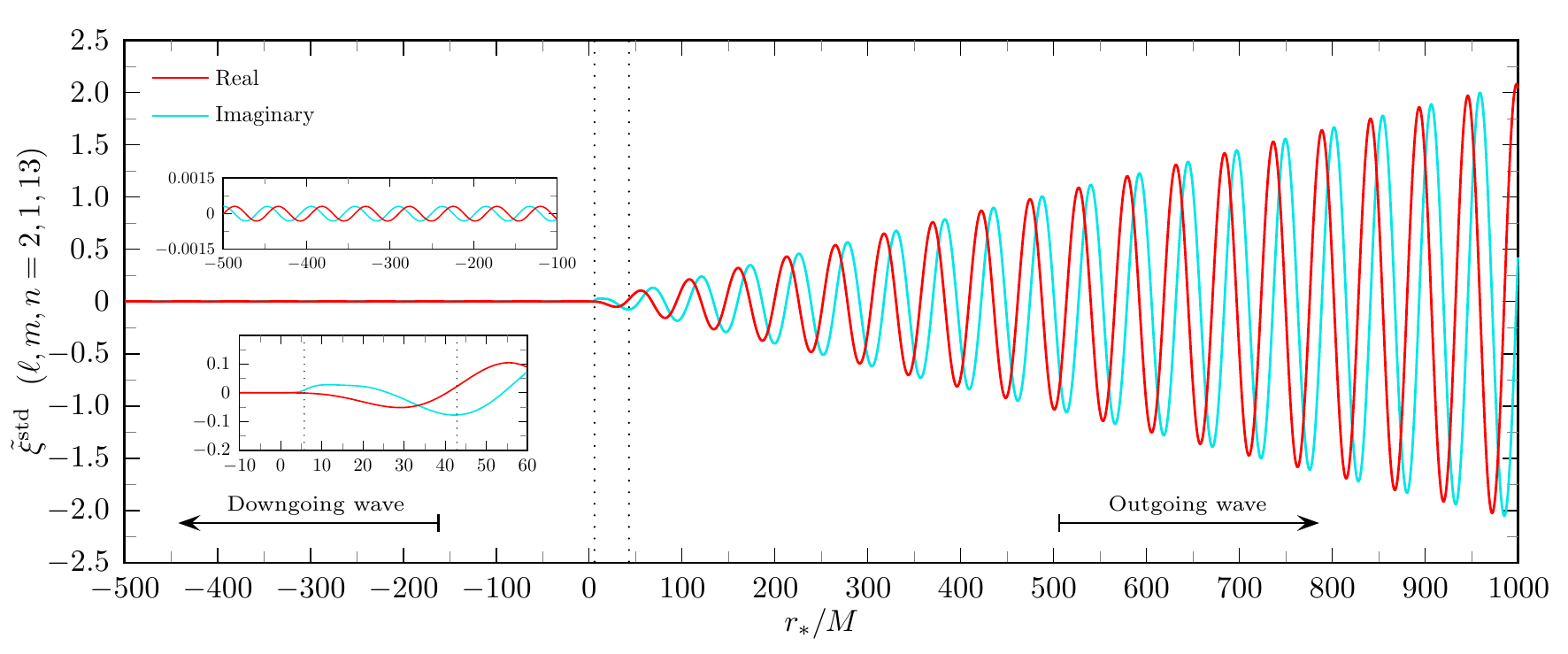}
\caption{
\label{fig:xi_p_Std}
Causally correct solution to Eq.~(\ref{eq:xiFD}),
$\tilde \xi^{\text{std}}$.
Dotted lines indicate the source libration region.}
\end{figure}
\end{enumerate}

The function $\tilde \xi^{\text{std}}_{p} 
= \tilde \xi^{H}_{p} - \la^{-} \tilde \xi^{-}_{h}
= \tilde \xi^{\infty}_{p} - \k^{+} \tilde \xi^{+}_{h}$ represents the 
standard solution to Eq.~(\ref{eq:xiFD}).  If the TD source were 
differentiable, we would be able to find the corresponding TD solution via 
an exponentially converging Fourier synthesis.  However, the source in this 
case is non-differentiable and we need an EHS-like trick to complete the
method.

\subsubsection{Restoring exponential convergence with extended particular 
solutions}

\begin{figure}[h!]
\includegraphics[scale=1]{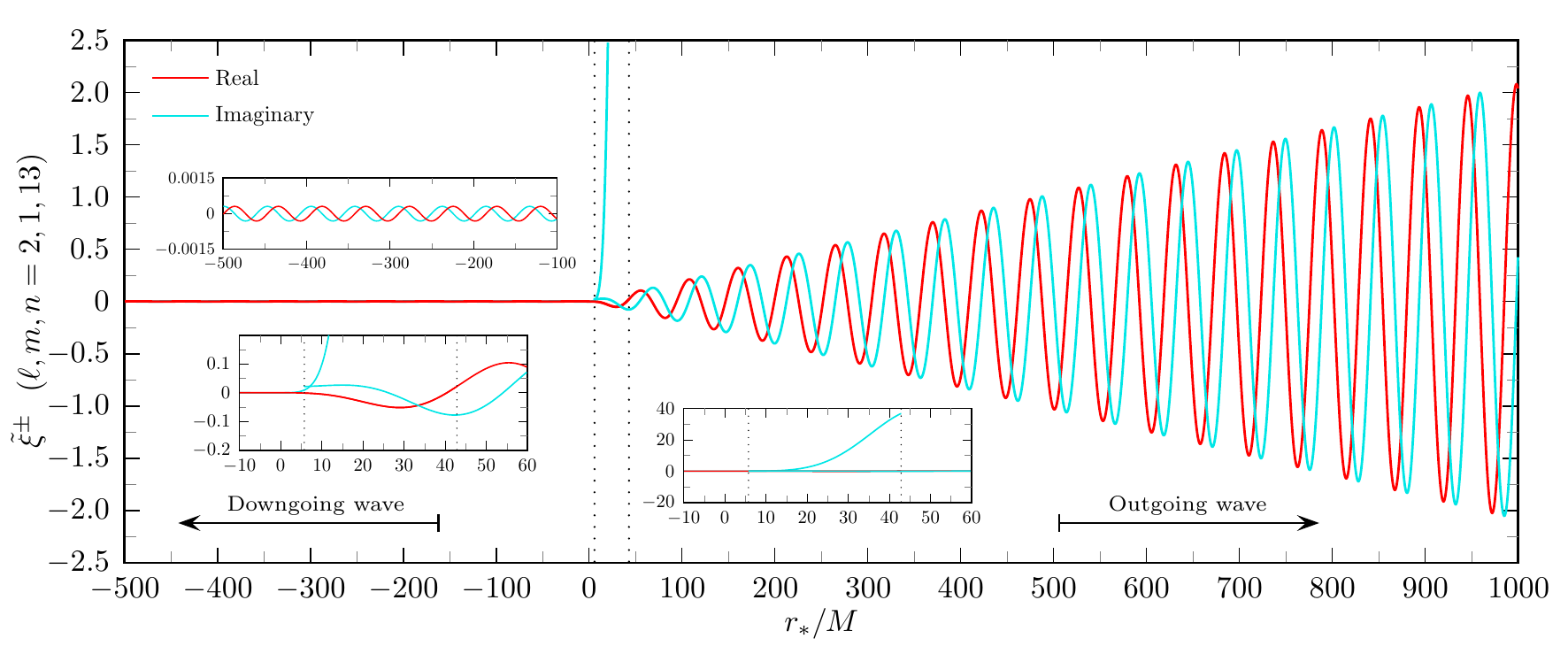}
\caption{
\label{fig:xi_EPS}
Causally correct extended particular solution,
$\tilde \xi^{\pm}$.  Note 
the difference within the libration region (shown in dotted lines), 
between $\tilde \xi^{\pm}$ and 
$\tilde \xi^{\text{std}}$ (Fig.~\ref{fig:xi_p_Std}).
The difference in TD convergence between these two solutions is
shown later in Fig.~\ref{fig:convVsX}.}
\end{figure}

The EHS of the Regge-Wheeler equation Eq.~(\ref{eq:RWeq}) are found by 
taking the constants $C^{\pm}_{\text{RW}}$ and scaling the unit-normalized 
homogeneous solutions 
\be
R^{\pm}_{\text{RW}} (r) \equiv C^{\pm}_{\text{RW}} \hat R^{\pm} (r) .
\label{eq:R_EHS}
\ee
These solutions are defined for all $r>2M$.  In like fashion we seek to find 
FD EPS of Eq.~(\ref{eq:xiFD}) and denote these by $\tilde \xi^{\pm}$.  We 
first find $\tilde \xi^{\pm}_{p}$ by separately integrating 
Eq.~(\ref{eq:xiFD}) with the modified source terms $R^{\pm}_{\text{RW}}$.  
The solutions are each made to match the exterior behavior of 
$\tilde \xi^{\text{std}}_{p}$ by adding the correctly scaled homogeneous 
solutions found in Step \ref{resolveStep} above.  We then define
\be
\tilde \xi^{+} \equiv \tilde \xi_{p}^{+} - \k^{+} \tilde \xi_{h}^{+},
\q \q
\tilde \xi^{-} \equiv \tilde \xi_{p}^{-} - \la^{-} \tilde \xi_{h}^{-}.
\ee
See Fig.~\ref{fig:xi_EPS}, which contrasts Fig.~\ref{fig:xi_p_Std}
in the source region.

These FD EPS can be transferred to the TD via Fourier series
\be
\xi^{\pm} (t,r) \equiv \sum_{n} \tilde \xi^{\pm} (r) e^{- i \o t}.
\ee
The solution to Eq.~(\ref{eq:xiExtended}) is then the weak solution,
\be
\xi_{o}^{\text{ext}} (t,r) =
 \xi^{+} (t,r)\, \th \left[ r - r_{p} (t) \right]
+
\xi^{-} (t,r)\, \th \left[ r_{p} (t) - r \right].
\label{eq:EPS}
\ee
The support for this claim has three legs.  Firstly, the same arguments about 
EHS, based on analytic continuation, made by Barack, Ori, and Sago in 
Ref.~\cite{BOS} appear to apply in extension to Eq.~(\ref{eq:xiExtended}) 
as well.  Secondly, we demonstrate existence \emph{numerically} by integrating 
the equation, with causal boundary conditions, and checking that the jump 
conditions (internal boundary conditions) at the particle are satisfied.  
One then appeals to the linearity of the equation to establish uniqueness.  
Finally, we have an independent numerical solution found through the method 
of partial annihilators and given in Eq.~(\ref{eq:xiPartAnn}).  We have 
confirmed that the two methods give entirely consistent solutions.  These 
results are covered in detail in the next section.

\begin{figure}[h!]
\includegraphics[scale=1]{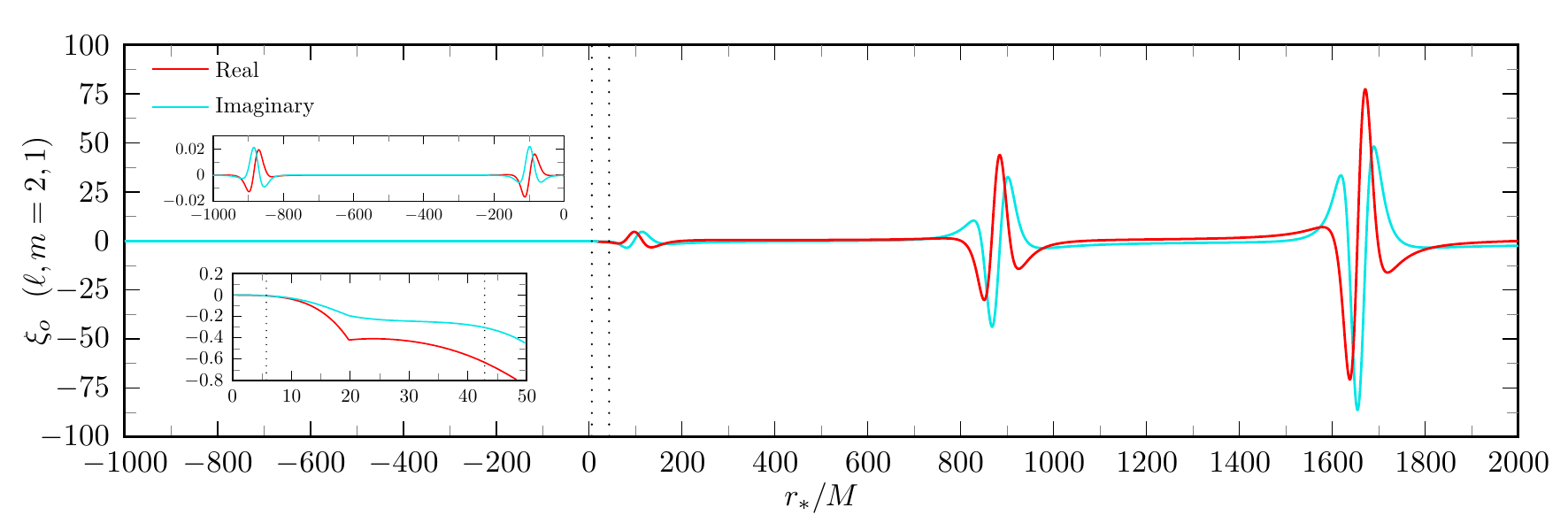}
\caption{
\label{fig:xiOdd}
The $\ell = 2, m = 1$ mode of the odd-parity RW-to-Lorenz gauge generator 
amplitude $\xi_{o}$.  The Lorenz gauge MP amplitude $h_{2}$ differs from 
$\xi_{o}$ only by a factor of $-2$.  Dotted lines indicate the region of 
libration.  Orbital parameters are given in the text.  While the field 
$h_{2}$ grows asymptotically, upon transforming to an orthonormal frame it 
would contribute a term that falls off as $1/r$.}
\end{figure}

\section{Results}
\label{sec:results}

\begin{figure}[h!]
\includegraphics[scale=1]{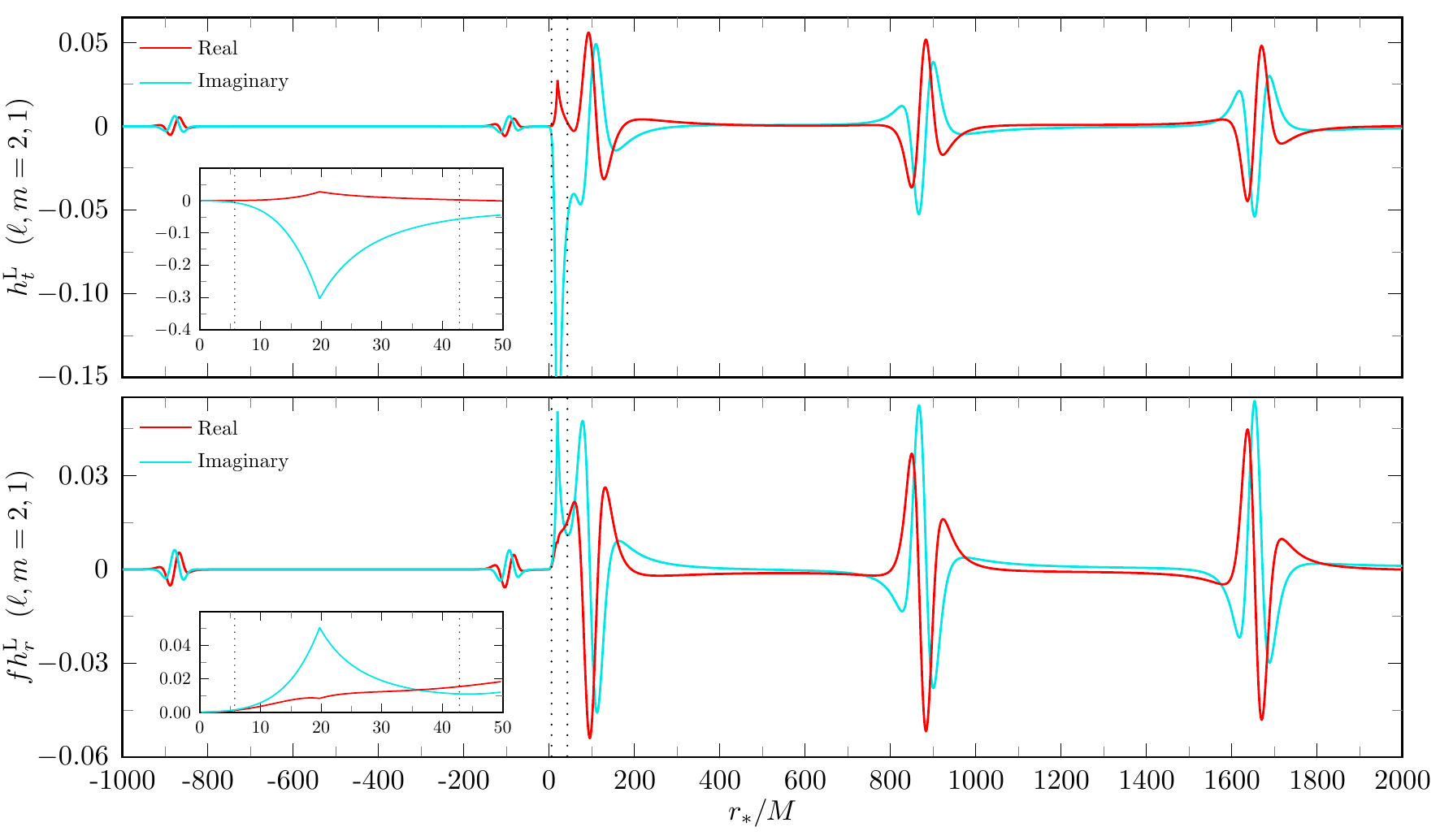}
\caption{
\label{fig:hthrL}
The $\ell = 2, m = 1$ mode of
the Lorenz gauge MP amplitudes $h^{\text{L}}_{t}$
and $h^{\text{L}}_{r}$.
Dotted lines indicate the region of libration.
Orbital parameters are given in the text.
Note (comparing to Fig.~\ref{fig:hthrRW}) 
the discontinuity at the location of the particle has vanished 
and the wave no longer grows asymptotically.
We plot $f h^{\text{L}}_{r}$ so we can see the wave behavior
near the horizon.}
\end{figure}

The methods of the previous section allow us to transform odd-parity 
solutions of the first-order Einstein equations from RW to Lorenz gauge.  
As an example, we consider an orbit with eccentricity $e = 0.764124$ and 
semi-latus rectum $p = 8.75455$.  This orbit was used in 
Fig.~\ref{fig:hthrRW} where we showed the RW amplitudes $h_{t}$ and $h_{r}$ 
for $\ell = 2$ and $m = 1$ ($h_{2} = 0$).  The MPs can be evaluated at any 
time but we chose to display results at $t = 93.58$ (where $t = 0$ is at 
the periapsis).  The RW modes are discontinuous at $r=r_p(t)$ and lack 
asymptotic flatness.  The gauge generator to go from RW to Lorenz gauge 
is computed for this same orbit and at the same time in the TD.  It is 
used to obtain the MPs in Lorenz gauge using Eq.~(\ref{eq:push}).
Fig.~\ref{fig:xiOdd} shows the $\ell=2$, $m=1$ amplitude of the gauge 
generator itself, which differs from $h_{2}$ in Lorenz gauge only by a factor 
of $-2$.  Fig.~\ref{fig:hthrL} shows the Lorenz gauge metric amplitudes 
$h^{\text{L}}_{t}$ and $h^{\text{L}}_{r}$ for the same mode.  The MPs are 
now $C^0$ at $r=r_p(t)$ and are asymptotically flat.

\begin{figure}
\includegraphics[scale=1]{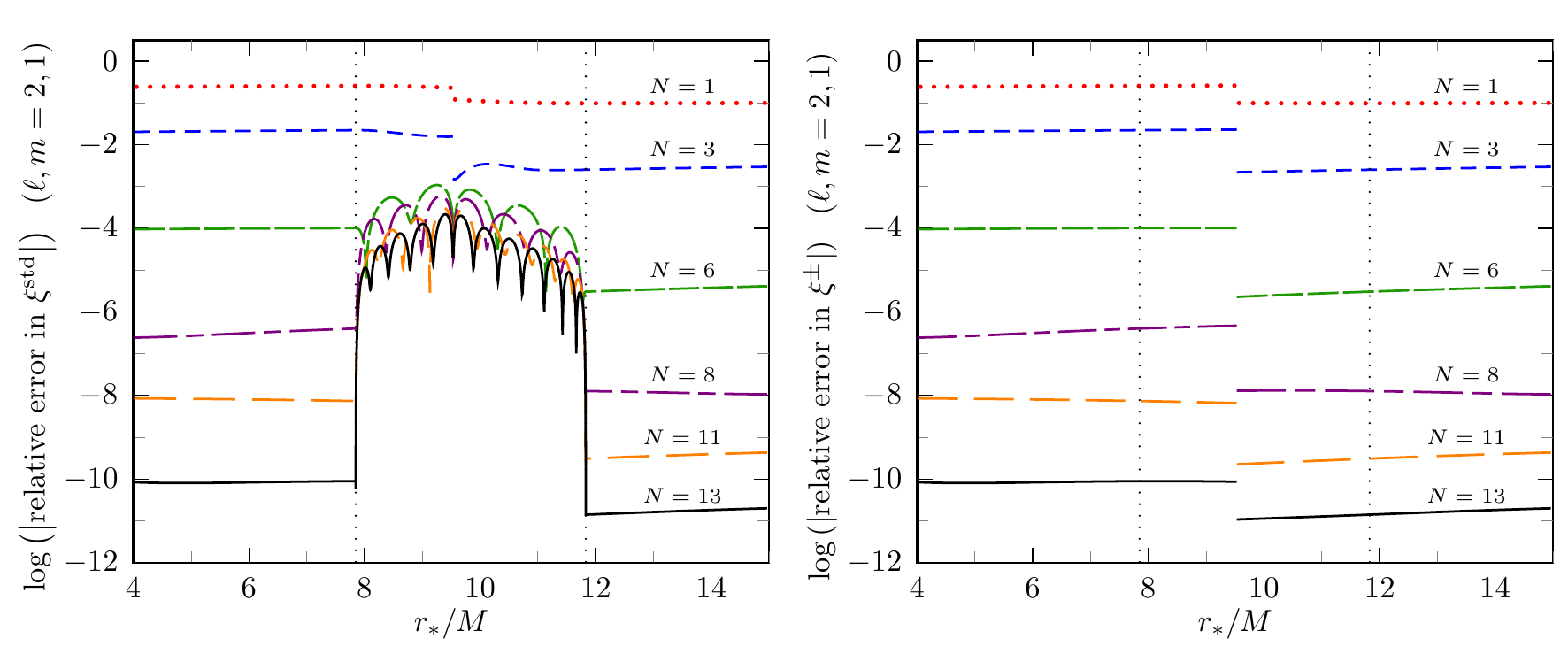}
\caption{
\label{fig:convVsX}
Self convergence of the $\ell = 2$, $m = 1$ mode of $\xi_{o}$.  We show 
results for the standard method in the left panel and our (equivalent)
partial annihilators and EPS on the right.  The orbital parameters are 
$e = 0.188917$ and $p = 7.50478$, and the fields are calculated at 
time $t = 96.44$.}
\end{figure}

Of key importance to our method is the exponential convergence of the TD 
solutions.  We can first consider self-convergence of the modes for all 
$r$.  As an example we choose an orbit with $e = 0.188917$ and 
$p = 7.50478$ at time $t = 96.44$.  In Fig.~\ref{fig:convVsX} we show the 
self-convergence of $\xi_{o}(r)$ for a set of partial Fourier sums over 
$n$ from $-N \le n \le +N$ for various $N$.  The right panel of this figure 
shows exponential self-convergence of the EPS method as a function of $r$, 
including at the particle.  This result is in contrast with the left panel 
which shows that the standard method is only algebraically convergent in 
the source libration region.  Note that the convergence is initially 
exponential before becoming algebraic around an error level of $10^{-4}$.  
This transition is due to the equation for $\xi_{o}$ having singular and
extended parts (see Eqs.~(\ref{eq:xiSing}) and (\ref{eq:xiExtended})).  
We find the singular part using EHS, which converges exponentially.  This 
part of the solution dominates the self-convergence in the left panel at 
first.  Eventually, the lack of differentiability of the extended source 
and the use (for comparison) of the standard Fourier series for that part 
of the field manifests itself.  The appearance of Gibbs behavior stalls 
the convergence in the libration region.

\begin{figure}
\includegraphics[scale=1]{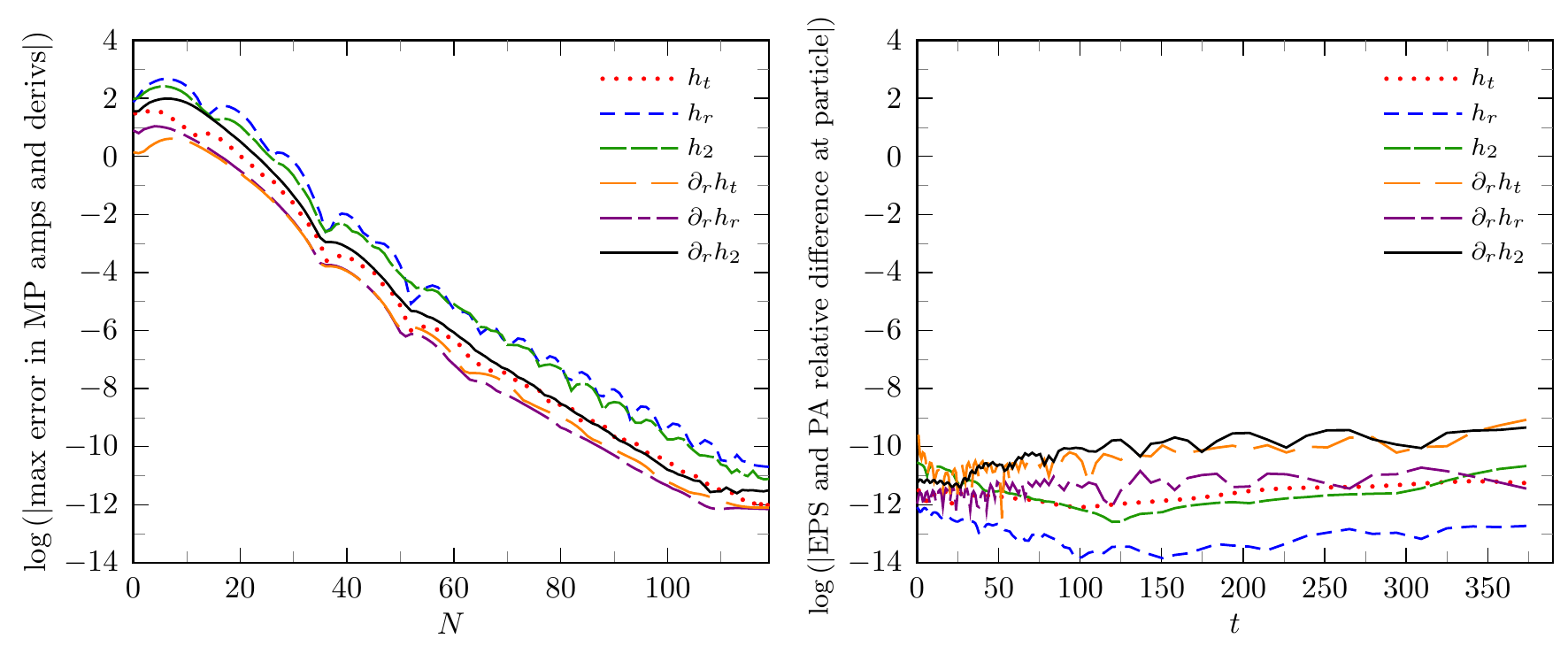}
\caption{
\label{fig:convVsN}
Convergence of the jump conditions as a measure of solution error.  In the 
left panel the partial annihilator method was used to compute the gauge 
generator and MPs for a high eccentricity orbit with $e = 0.764124$ and 
$p = 8.75455$.  Partial Fourier sums over $n$ are computed with 
$-N \le n \le +N$ and for various $N$.  Exponential convergence is exhibited 
in the various sums.  See text for further discussion.  In the right panel 
a comparison is made of discrepancies between the EPS method and the PA 
method as a function of time about the orbit.  See further discussion in the 
text.}
\end{figure}

Beyond self-convergence, we can check absolute convergence to the 
analytically-known jump conditions.  This test can be applied to both the 
MP amplitudes and their first radial derivatives.  As shown in the left 
panel of Fig.~\ref{fig:convVsN}, we find exponential convergence to the 
analytically computed values in Eq.~(\ref{eq:LGJumps}), in this case using 
the partial annihilator method.  Here the orbit is the more eccentric one 
with $e = 0.764124$ and $p = 8.75455$.  Each partial Fourier sum ranges 
over all harmonics from $-N \le n \le +N$ for different values of $N$ as 
seen on the horizontal axis.  The jump conditions are time dependent and 
thus we compare our results at several moments in time (in this case at 
20 points) throughout the orbit.  The left panel plots the maximum error 
encountered in each quantity throughout an orbit.  Since the Lorenz gauge 
amplitudes are all $C^{0}$, we plot absolute convergence for the jumps in 
the amplitudes themselves (which are expected to converge to 0) and relative 
convergence for the jumps in the $r$ derivatives of the amplitudes.  The 
convergence appears to bottom out around $10^{-12}$ to $10^{-11}$.

In the right panel of Fig.~\ref{fig:convVsN} we compare the accuracy of 
the EPS and PA methods.  For the jumps in Lorenz gauge MP amplitudes and 
their radial derivatives we show the relative error between the two methods 
as a function of time throughout one orbit.  The same high eccentricity 
orbit is used, though to compare the two methods the partial Fourier 
sums were fixed and taken to range over $-85 \le n \le 106$.  The two 
methods agree with each other to the level of $10^{-12}$ to $10^{-10}$.

Although we have only displayed results in this paper for the 
$\ell = 2, m = 1$ mode, we have run the code on many different modes and 
for different orbits.  We have no difficulty in computing the gauge 
transformation from RW to Lorenz for odd-parity modes with high accuracy.  
It now remains for us to apply these methods to the even-parity part of 
the gauge transformation, a somewhat more involved procedure.  We will 
turn to that issue in a subsequent paper.  

\section{Conclusion}

This paper is the first of two on the transformation of metric perturbations 
from Regge-Wheeler gauge to Lorenz gauge.  This first paper was confined 
to treating the odd-parity part of the MPs and devoted much of the 
discussion to the development of two new analytic/numerical methods for 
using frequency domain methods to find accurate solutions in the time 
domain.  The follow-on paper will be primarily devoted to discussing the 
analytic problem of finding the even-parity part of the gauge transformation, 
and will draw upon the numerical methods which we have detailed here.

\acknowledgments

The authors thank Chad Galley and Ian Hinder for helpful discussions.  We 
also appreciate suggestions made by the referee.  CRE acknowledges support 
from the Bahnson Fund at the University of North Carolina--Chapel Hill.

\appendix

\section{Asymptotic expansions and boundary conditions}

\label{sec:asympExp}

In the RWZ formalism it is useful to compute asymptotic expansions of the 
master functions about $r=\infty$ to provide boundary conditions for starting
numerical integrations at finite radius.  In this paper, the inhomogeneous 
equation for the gauge generator, Eq.~(\ref{eq:xiFD}), has a source term 
that is non-compact.  This fact leads to an inhomogeneous recurrence
relation for the asymptotic expansion of $\tilde \xi_{o}$ that requires as
input the asymptotic expansion of the source term.

We start by writing
\be
\label{eq:JostOdd}
\tilde \xi_{o} = r J_{o}(r) e^{i \o r_{*}},
\ee  
where $J_{o}(r)$ is the Jost function~\cite{Chandra_1983}, which goes
to 1 at infinity.  We use Eq.~(\ref{eq:CPMRW}) to express the RHS of 
Eq.~(\ref{eq:xiFD}) in terms of the CPM function.
Then we Fourier transform that function 
and plug in Eq.~(\ref{eq:JostOdd}) to obtain
\begin{align}
r f \frac{d^{2}}{dr^{2}} J_{o}
+
2 \l 1+ i \o r  - \frac{M}{r} \r \frac{d}{dr} J_{o}
+ \l 2 i \o + \frac{2M}{r^{2}}  - \frac{\ell (\ell + 1)}{r} \r
 J_{o}
= - i \o J_{R}.
\label{eq:JostXi}
\end{align}
Here $J_{R} = J^{+}_{\ell m n}$ from App.~D of \cite{HE_2010}.
Now, we assume the following forms of $J_{o}$, and $J_{R}$,
\be
J_{o} (r) = \sum_{j=0}^{\infty} \frac{a^{o}_{j}}{(r \o)^{j}},
\q \q
J_{R} (r) = \sum_{j=0}^{\infty} \frac{a^{R}_{j}}{(r \o)^{j}}.
\ee
Plugging these in and assuming the equation is satisfied order-by-order
gives the coupled recurrence formula,
\be
 2 i (j-1)  a^{o}_{j}
= \Big[ (j-2) (j-1)
- \ell (\ell + 1) \Big]  a^{o}_{j-1}
 + 2 M \o \Big[ 1-(j-2)^{2} \Big]
a^{o}_{j-2}
+ i a^{R}_{j}.
\ee
The coefficients $a^{R}_{j} = a_{j}$, given in Eq.~D5 of \cite{HE_2010}.
Assuming $a_{j}^{o} = 0$ for $j < 0$, this recurrence allows for the 
calculation of all $a_{j}^{o}$.
Note that this recurrence fails at $j=1$, which represents the homogeneous
solution to Eq.~(\ref{eq:JostXi}).  We can choose that coefficient to be
anything.

The particular solution here
is identical to the homogeneous solution to the fourth-order 
equation, given asymptotically in Eq.~(\ref{eq:homog4th}).  
We can use this asymptotic expansion for both situations.

On the horizon side, where the potential falls away exponentially,
it is enough to use the expression in Eq.~(\ref{eq:xiCMinus})
and a sufficiently large and negative $r_{*}$ starting location 
for integration.  A Taylor expansion could be used if the starting location 
were farther from the horizon.  The boundary conditions to the second-order
homogeneous solutions are exactly analogous to those given in the 
odd-parity recurrence of App.~D in \cite{HE_2010}.  The only difference is 
a change of the spin parameter in the potential from 2 to 1.

\bibliography{GaugePaperI}

\end{document}